\pdfoutput=1
\documentclass[letterpaper]{article}
\usepackage[dvipdfmx]{graphicx} 
\usepackage{bmpsize}
\usepackage{aaai20}
\usepackage{times}
\usepackage{helvet}
\usepackage{multirow}
\usepackage{courier}
\usepackage{graphicx}
\usepackage{booktabs}
\usepackage[toc,page]{appendix}
\usepackage{url}

\usepackage{subcaption}
\usepackage{xcolor}
\usepackage{soul}

\usepackage{todonotes}

\newcommand{\citet}[1]{\citeauthor{#1} \shortcite{#1}}
\newcommand{\citep}{\cite}

\begin{document}
%
\title{Man is to Person as Woman is to Location: \\Measuring Gender Bias in Named Entity Recognition}
\author{Ninareh Mehrabi, Thamme Gowda, Fred Morstatter, Nanyun Peng, Aram Galstyan, \\
 University of Southern California\\
        Information Sciences Institute\\
        \{ninarehm, tg, fredmors, npeng, galstyan\}@isi.edu}
\maketitle
\begin{abstract}
\begin{quote}

We study the bias in several state-of-the-art named entity recognition (NER) models---specifically, a  difference in the ability to recognize male and female names as PERSON entity types.
We evaluate NER models on a dataset containing  139 years of U.S. census baby names and find that relatively more female names, as opposed to male names, are not recognized as PERSON entities. We study the extent of this bias in several NER systems that are used prominently in industry and academia. In addition, we also report a bias in the datasets on which these models were trained. 
The result of this analysis yields a new benchmark for gender bias evaluation in named entity recognition systems.
The data and code for the application of this benchmark will be publicly available for researchers to use.\footnote{\label{projecturl}\url{https://github.com/Ninarehm/NERGenderBias}}

\end{quote}
\end{abstract}
\section{Introduction}
Machine learning and AI systems are becoming omnipresent in everyday lives. Recently, attention has been directed to problems concerning fairness and algorithmic bias. Some progress has been made on the analysis of gender stereotyping in different natural language processing (NLP) components, such as word embedding \cite{bolukbasi2016man,zhao2018learning}, co-reference resolution \cite{zhao2018gender}, machine translation \cite{font2019equalizing} and sentence encoders \cite{may2019measuring}. In this work we study bias in named entity recognition (NER) systems and show how they can propagate gender bias by analyzing the 139-year history of U.S. male and female names from census data.

\begin{figure}
\includegraphics[width=0.45\textwidth,height=0.37\textwidth,trim=4cm 9cm 4cm 9cm,clip=true]{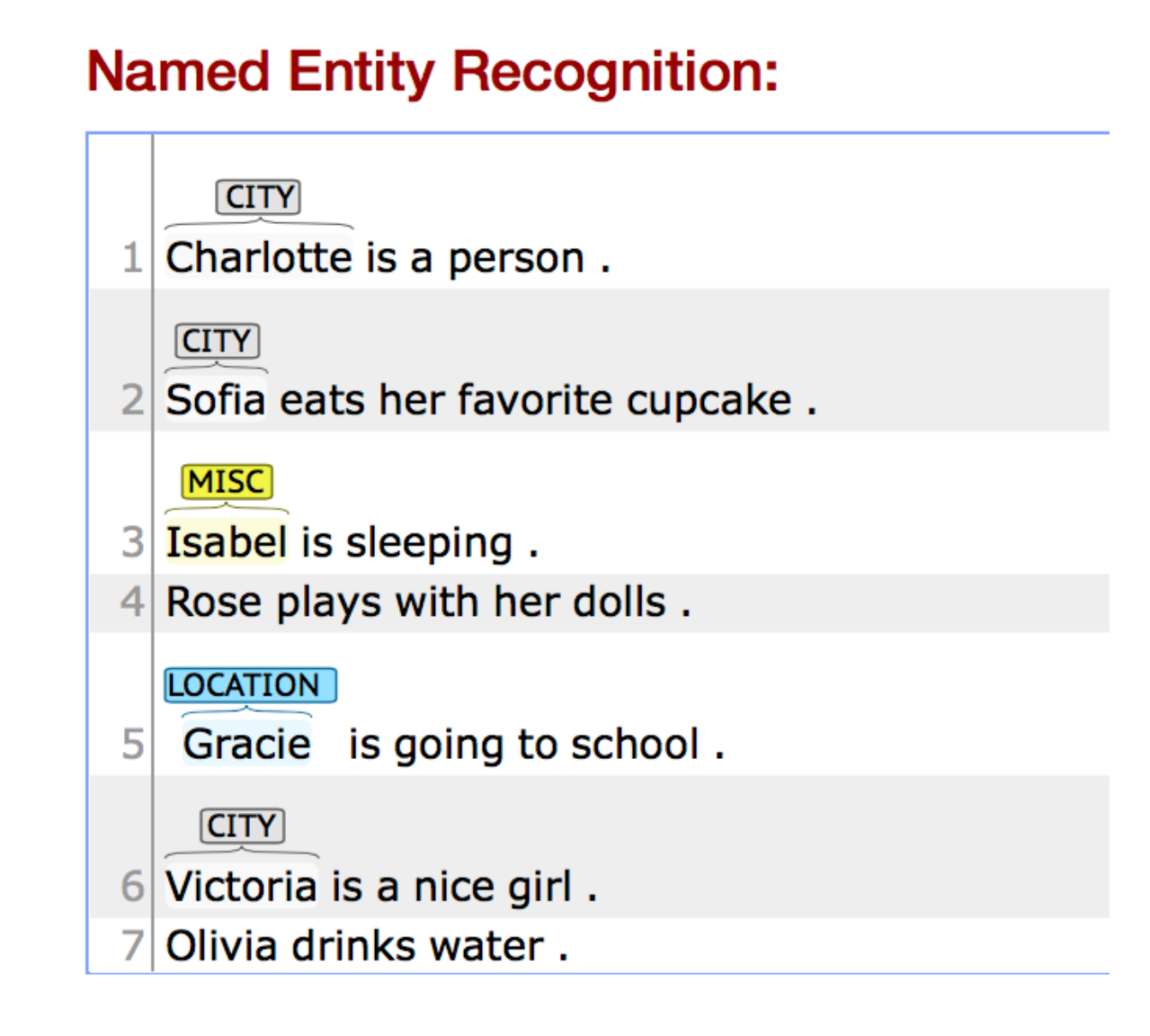}
\caption{Examples of PERSON entities that are wrongfully tagged as non-PERSON or NULL entities by CoreNLP.}
\label{motivation}
\end{figure}

Our experiments show that widely used named entity recognition systems are susceptible to gender bias. We find that relatively more female names were tagged as non-\textsc{PERSON} than male names even though the names were used in a context where they should have been marked as \textsc{PERSON}. An example is ``Charlotte,'' ranked as the top 8th most popular female U.S. baby name in 2018. ``Charlotte'' is almost always tagged wrongfully as a location by the state-of-the-art NER systems despite being used in a context when it is clear that the entity should be a person. Figure \ref{motivation} has more examples with names that are either not recognized as an entity or wrongfully tagged.
We show that there are many instances of such cases throughout history in the real world, and that there are more female names than male names that are incorrectly tagged. Moreover, based on this same U.S. census data, we find that this miscategorization affects more women than men. This serves as our definition of bias which considers the differences between gender groups following the statistical parity notion of fairness \cite{NIPS2017_6995,Dwork:2012:FTA:2090236.2090255}.

The contributions of this paper are fourfold:
\begin{enumerate}
    \item We introduced a benchmark dataset
    that tests NER models for gender bias. This benchmark can be run on any blackbox system.
    \item We studied the existence and extent of gender bias in current NER systems by analysing a 139-year history of names according to the statistical parity notion of fairness \cite{NIPS2017_6995,Dwork:2012:FTA:2090236.2090255}.
    \item We compared different versions of models and found that newer versions are amplifying gender bias in an effort to boost performance. Based on our observations, we defined a new source of bias that arises from version updates in the systems. 
    \item Finally, we analyzed datasets currently used for training many NER models and found the extent of gender bias in these datasets.
\end{enumerate}

\section{Models and Experiments}
To measure the existence of bias in NER systems, we evaluated five named entity models used in research and industry. We used Flair~\cite{akbik2018coling,akbik-etal-2019-flair}, CoreNLP version 3.9~\cite{manning-EtAl:2014:P14-5,Finkel:2005:INI:1219840.1219885}, and Spacy version 2.1 with small, medium, and large models.\footnote{\small{\url{https://spacy.io}}}
We test these models against 139 years of U.S. census data\footnote{\small{\url{http://www.ssa.gov/oact/babynames/names.zip}}} from years 1880 to 2018. Our benchmark evaluates these models based upon how well they recognize these names as a PERSON entity. 
\subsection{Benchmark}
Our benchmark dataset consists of nine templates listed in Table \ref{Templates} which are templated sentences that start with the existing names in the census data followed by a sentence that represents a human-like activity. The aim was to test the performance of the NER from different perspectives. Template 1, containing just the name, purely tests the name itself and reveals something about the distribution of the training data. Template 4 is designed to direct the model to tag the name as a person. Template 3 may reveal more subtle gender bias that stems from society as, historically, men have received greater levels of education from women. It is possible that the error rate may be even higher for female names under Template 3.

\begin{figure*}[!bt]
\begin{subfigure}[b]{0.33\textwidth}
\caption{Error Type-1 Weighted}
\includegraphics[width=\textwidth,height=0.5\textwidth,trim=0cm 0cm 0cm 0cm,clip=true]{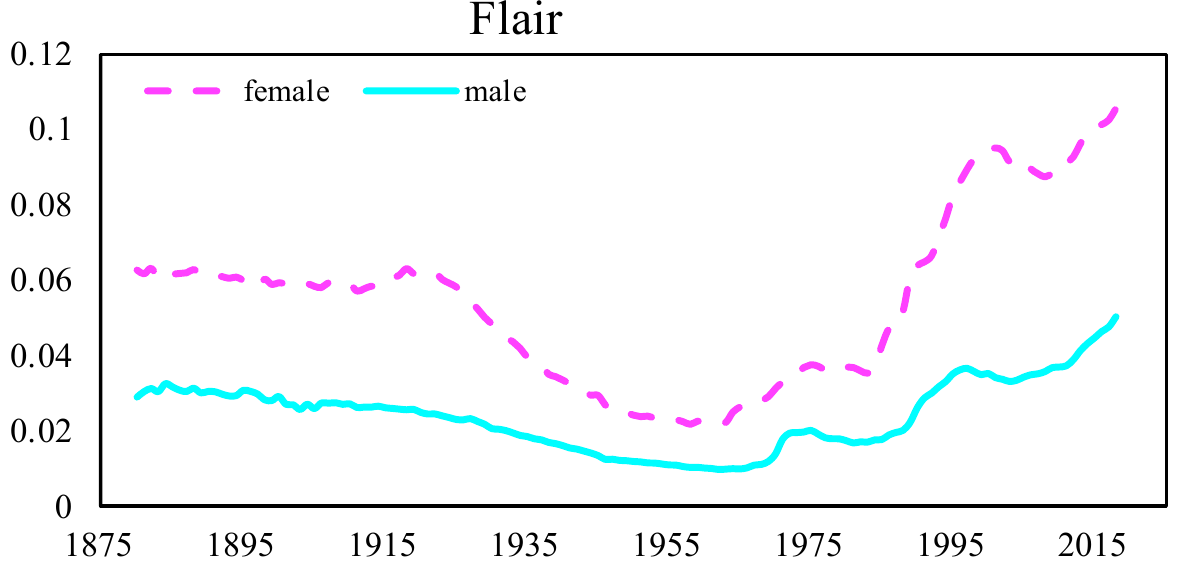}
\end{subfigure}
\begin{subfigure}[b]{0.33\textwidth}
\caption{Error Type-2 Weighted}
\includegraphics[width=\textwidth,height=0.5\textwidth,trim=0cm 0cm 0cm 0cm,clip=true]{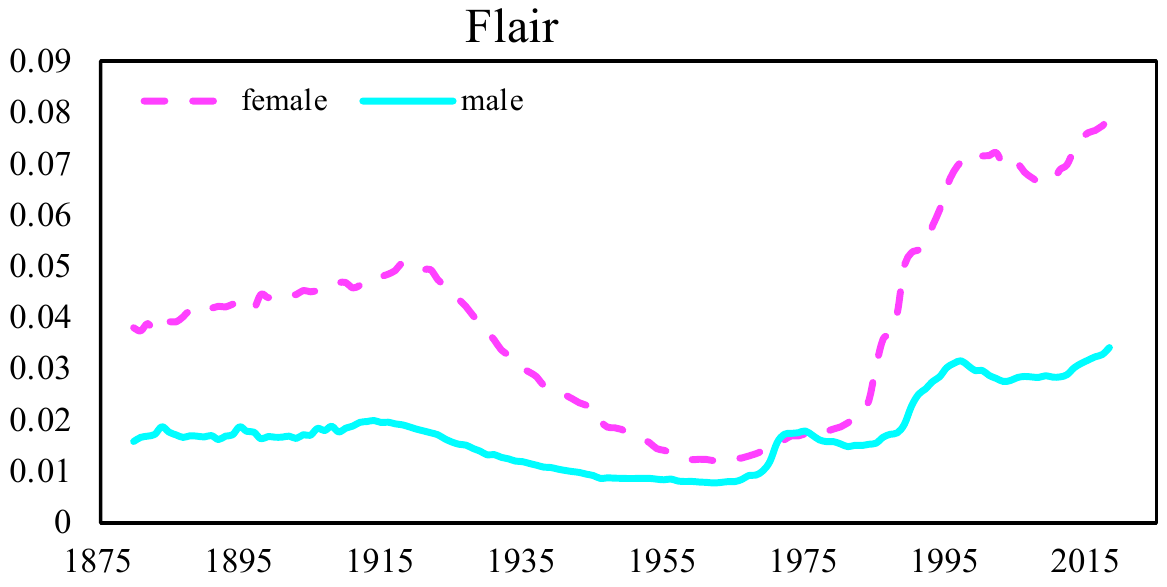}
\end{subfigure}
\begin{subfigure}[b]{0.33\textwidth}
\caption{Error Type-3 Weighted}
\includegraphics[width=\textwidth,height=0.5\textwidth,trim=0cm 0cm 0cm 0cm,clip=true]{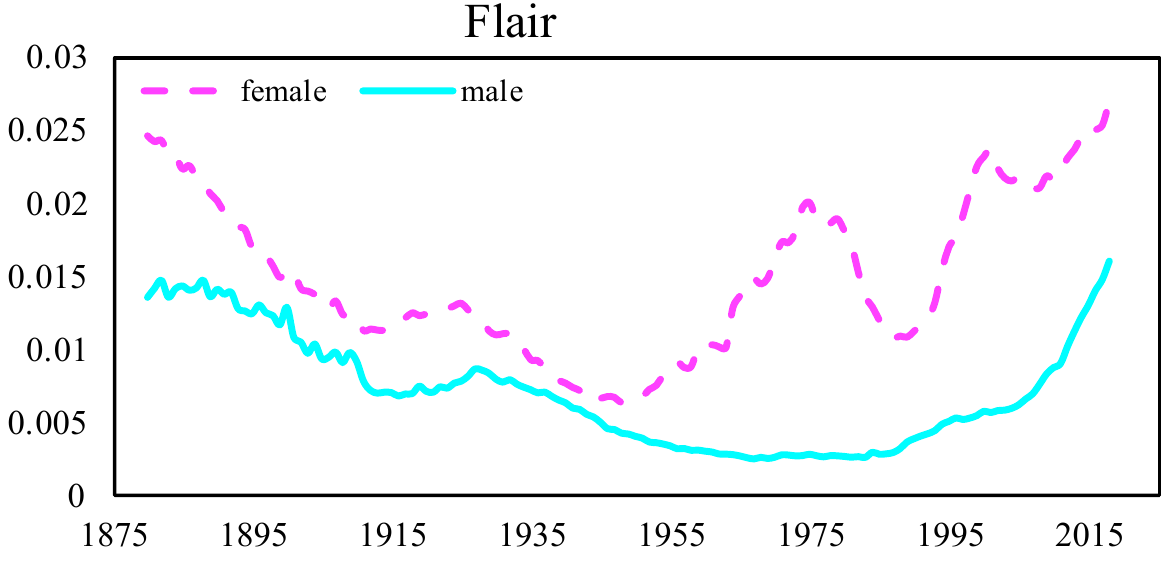}
\end{subfigure}
\begin{subfigure}[b]{0.33\textwidth}
\includegraphics[width=\textwidth,height=0.5\textwidth,trim=0cm 0cm 0cm 0cm,clip=true]{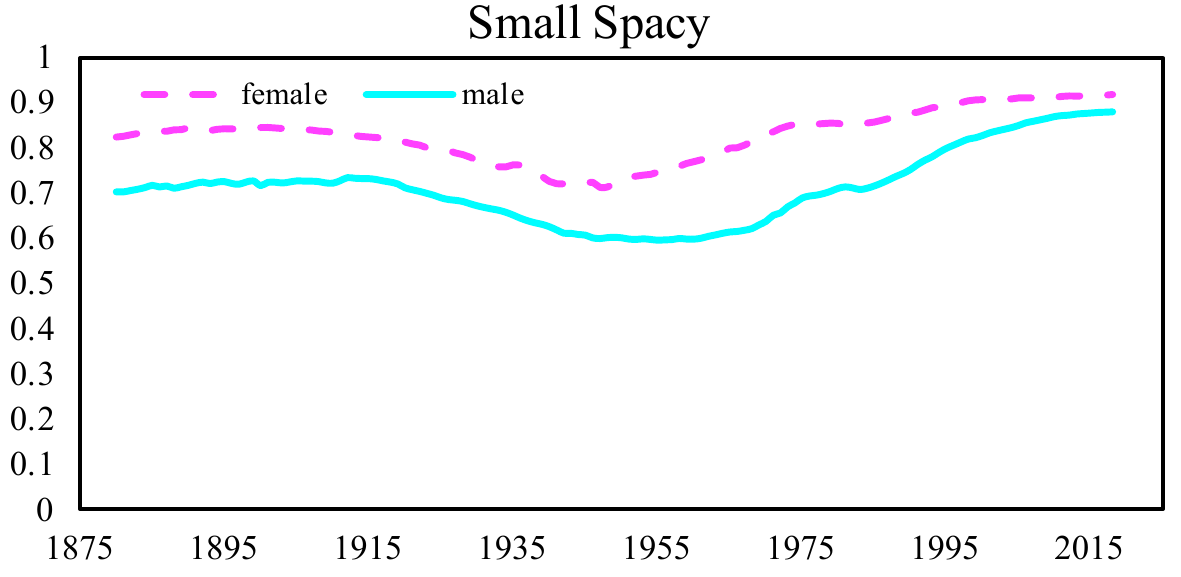}
\end{subfigure}
\begin{subfigure}[b]{0.33\textwidth}
\includegraphics[width=\textwidth,height=0.5\textwidth,trim=0cm 0cm 0cm 0cm,clip=true]{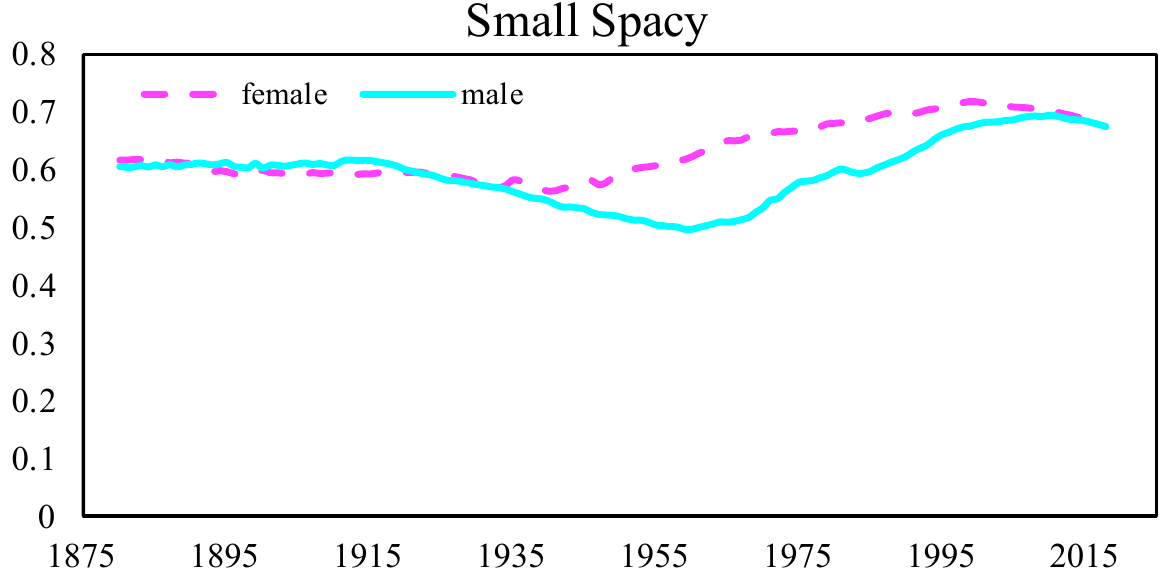}
\end{subfigure}
\begin{subfigure}[b]{0.33\textwidth}
\includegraphics[width=\textwidth,height=0.5\textwidth,trim=0cm 0cm 0cm 0cm,clip=true]{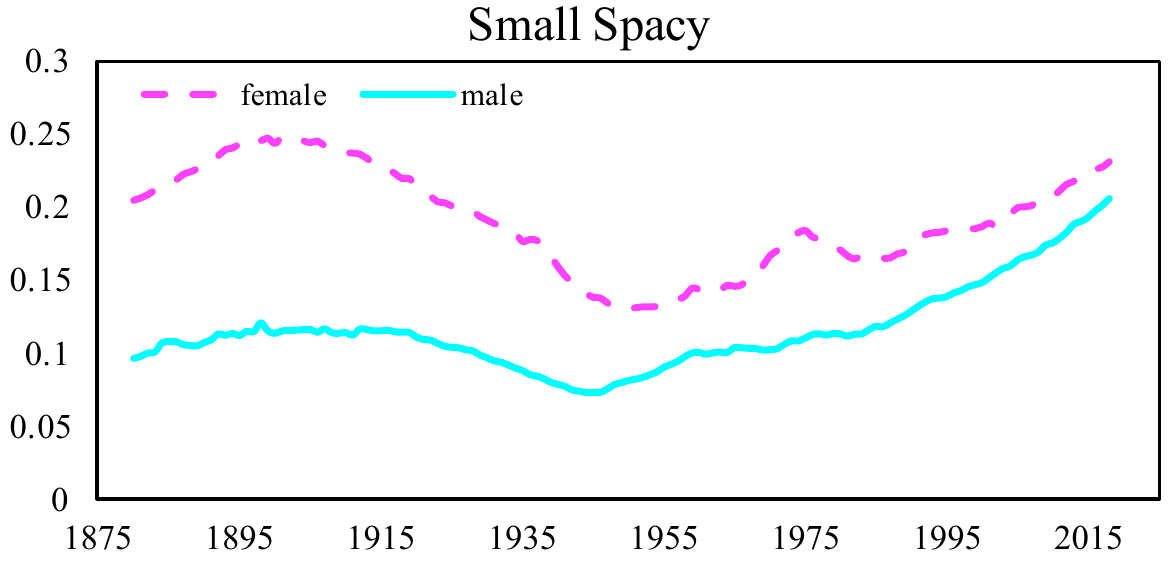}
\end{subfigure}
\begin{subfigure}[b]{0.33\textwidth}
\includegraphics[width=\textwidth,height=0.5\textwidth,trim=0cm 0cm 0cm 0cm,clip=true]{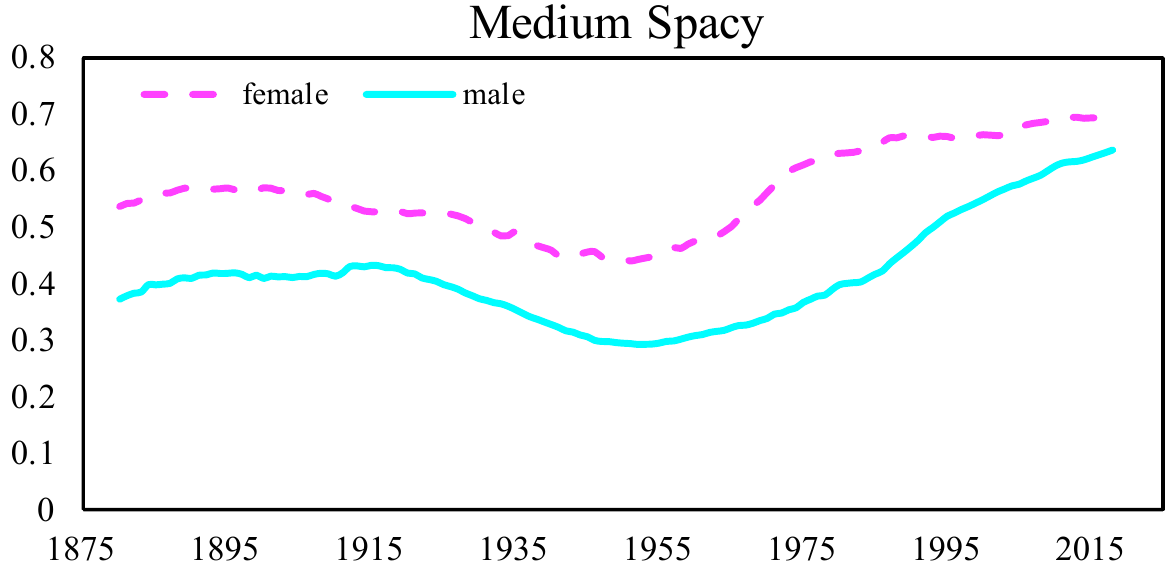}
\end{subfigure}
\begin{subfigure}[b]{0.33\textwidth}
\includegraphics[width=\textwidth,height=0.5\textwidth,trim=0cm 0cm 0cm 0cm,clip=true]{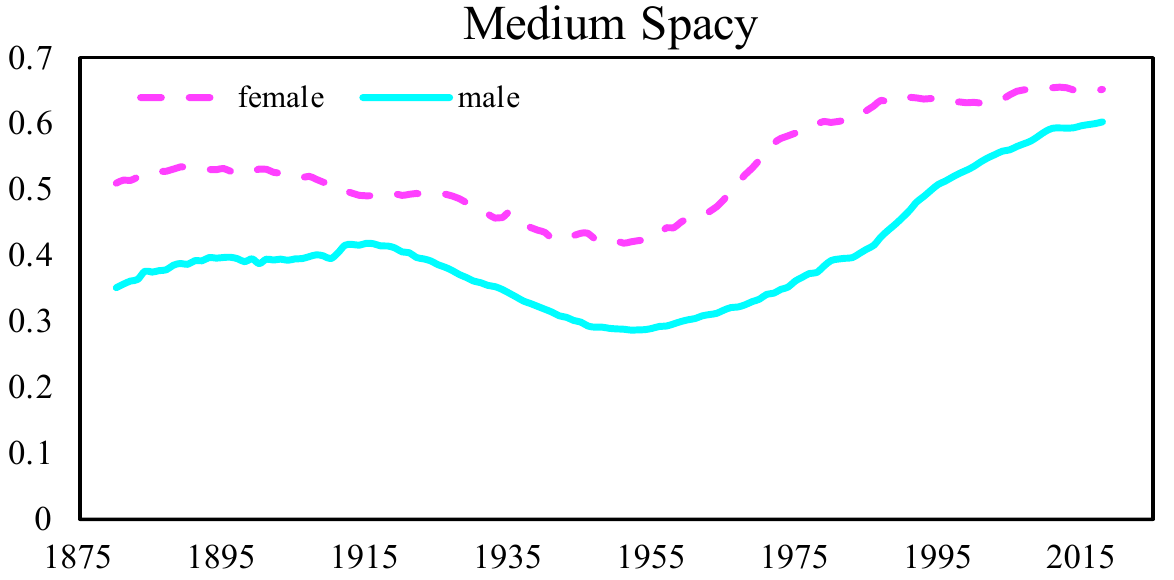}
\end{subfigure}
\begin{subfigure}[b]{0.33\textwidth}
\includegraphics[width=\textwidth,height=0.5\textwidth,trim=0cm 0cm 0cm 0cm,clip=true]{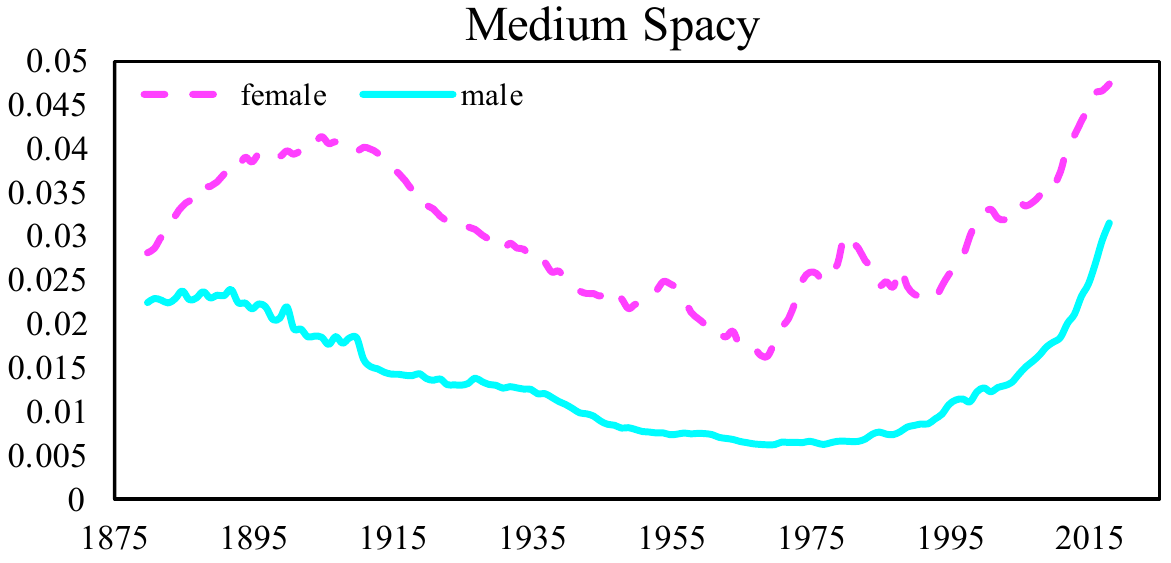}
\end{subfigure}
\begin{subfigure}[b]{0.33\textwidth}
\includegraphics[width=\textwidth,height=0.5\textwidth,trim=0cm 0cm 0cm 0cm,clip=true]{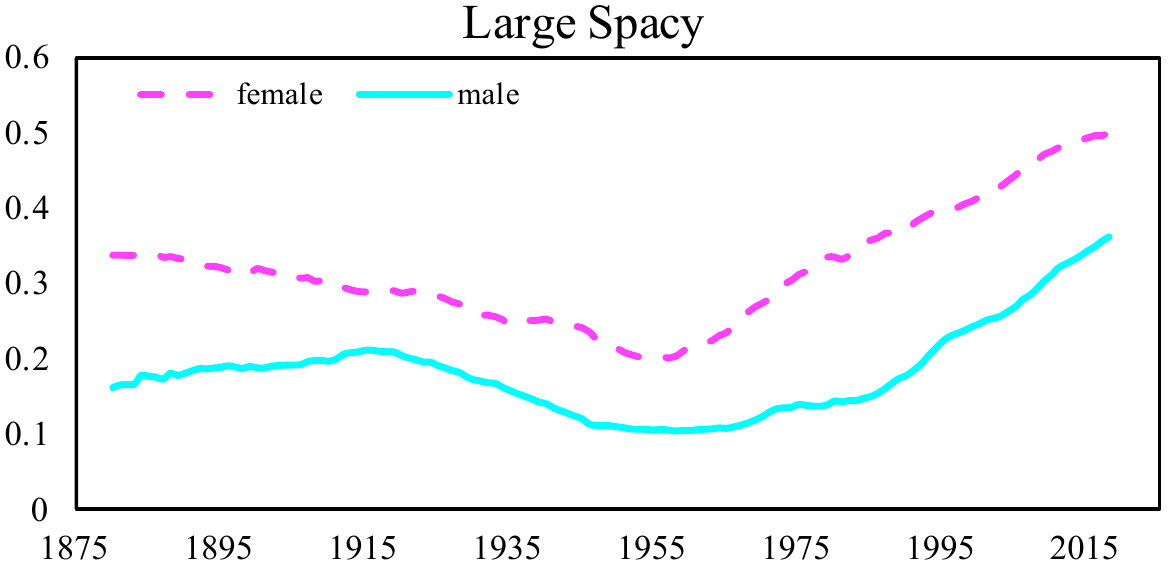}
\end{subfigure}
\begin{subfigure}[b]{0.33\textwidth}
\includegraphics[width=\textwidth,height=0.5\textwidth,trim=0cm 0cm 0cm 0cm,clip=true]{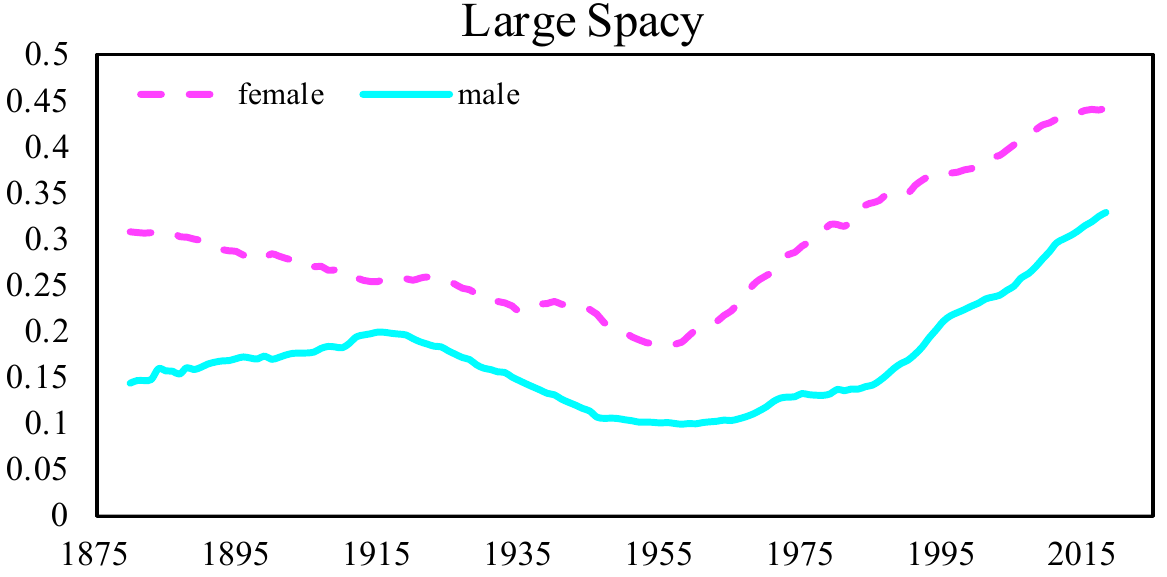}
\end{subfigure}
\begin{subfigure}[b]{0.33\textwidth}
\includegraphics[width=\textwidth,height=0.5\textwidth,trim=0cm 0cm 0cm 0cm,clip=true]{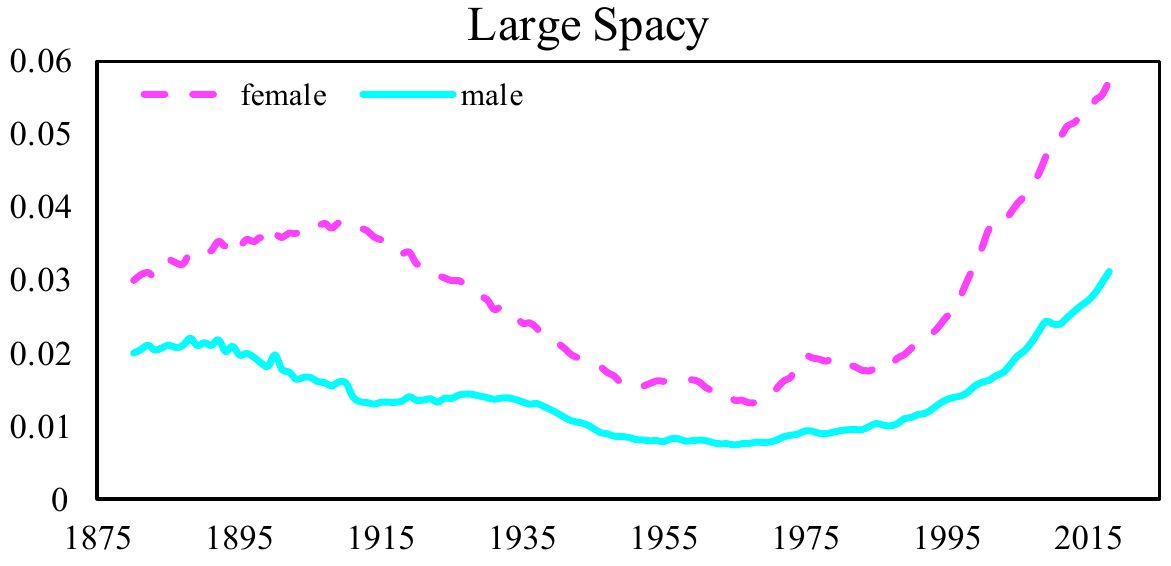}
\end{subfigure}
\begin{subfigure}[b]{0.33\textwidth}
\includegraphics[width=\textwidth,height=0.5\textwidth,trim=0cm 0cm 0cm 0cm,clip=true]{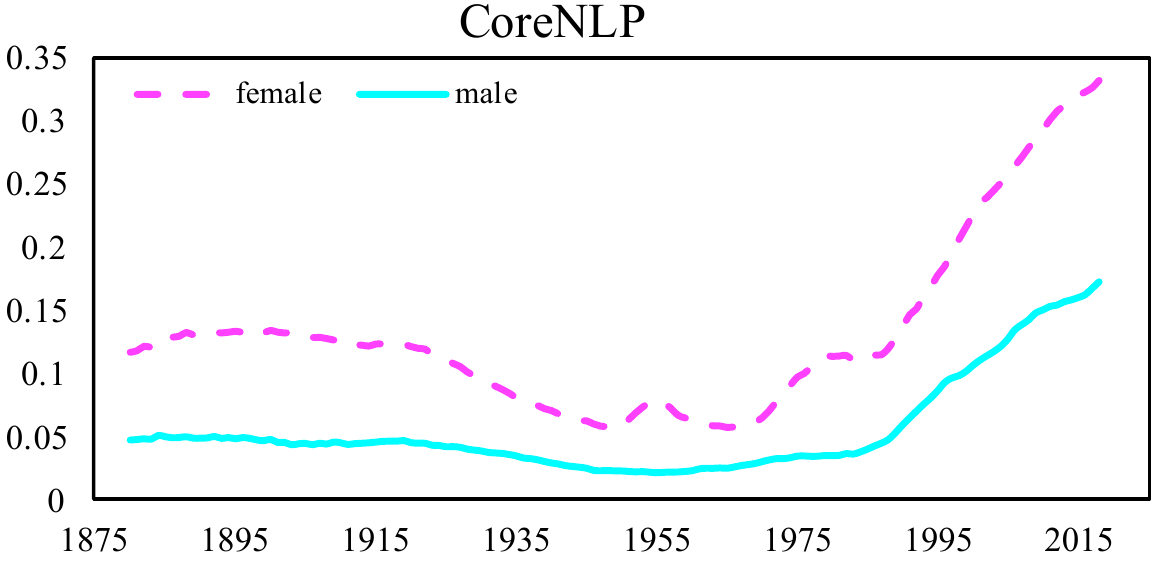}
\end{subfigure}
\begin{subfigure}[b]{0.33\textwidth}
\includegraphics[width=\textwidth,height=0.5\textwidth,trim=0cm 0cm 0cm 0cm,clip=true]{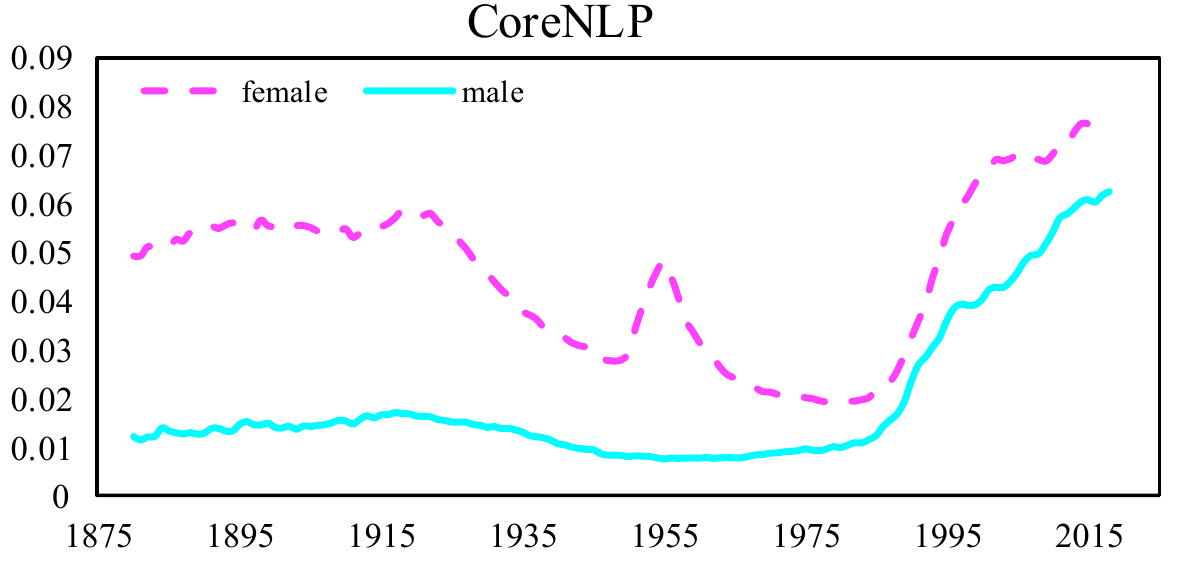}
\end{subfigure}
\begin{subfigure}[b]{0.33\textwidth}
\includegraphics[width=\textwidth,height=0.5\textwidth,trim=0cm 0cm 0cm 0cm,clip=true]{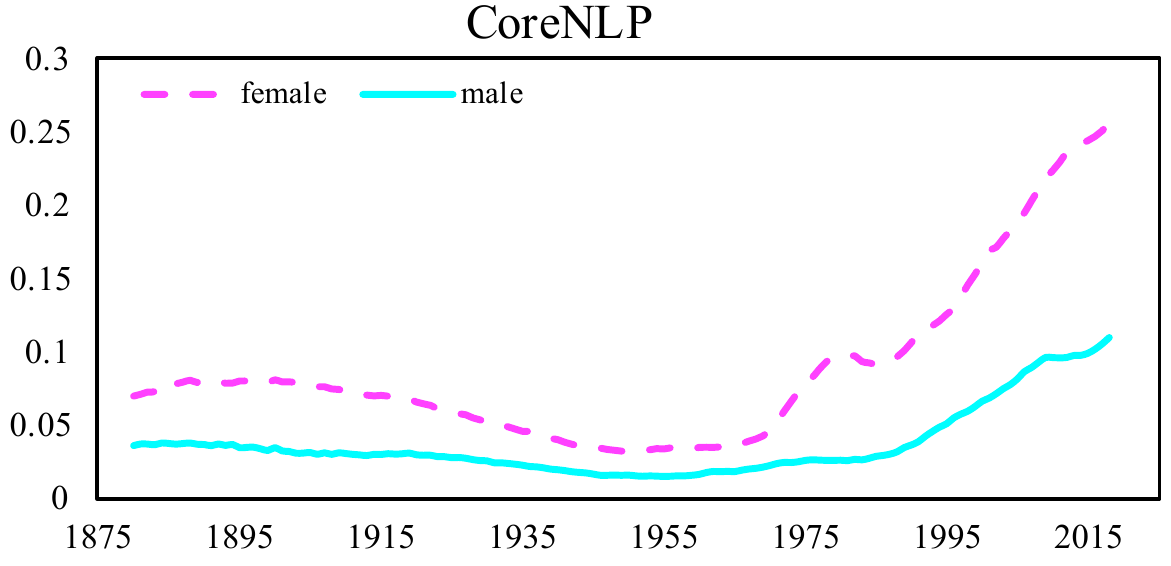}
\end{subfigure}
\caption{Results from different models that spanned the 139-year history of baby names from the census data on different error types for the weighted cases using template \#4. Female names have higher error rates for all the cases. The y axis shows the calculated error rates for each of the error types as described in their corresponding formulas, and the x axis represents the year in which the baby name was given.}
\label{result1}
\end{figure*}
\subsection{Experimental Design and Results}
\begin{table}
\centering
\begin{tabular}{ p{1cm}p{5.0cm}}
 \toprule
\#&Template Sentence\\
 \midrule
 1&\textbf{\textless Name\textgreater}\\[0.5pt]
 2&\textbf{\textless Name\textgreater} is going to school\\[0.5pt]
 3&\textbf{\textless Name\textgreater} is at school\\[0.5pt]
 4&\textbf{\textless Name\textgreater} is a person\\[0.5pt]
 5&\textbf{\textless Name\textgreater} is eating food\\[0.5pt]
 6&\textbf{\textless Name\textgreater} is going to grocery shop\\[0.5pt]
 7&\textbf{\textless Name\textgreater} is going to work\\[0.5pt]
 8&\textbf{\textless Name\textgreater} is a nurse\\[0.5pt]
 9&\textbf{\textless Name\textgreater} is a doctor\\[0.5pt]
 \bottomrule
\end{tabular}
\caption{Templates that form our benchmark with their corresponding numbers as referenced in the paper. Template 1 is the ``no context'' template.}
\label{Templates}
\end{table}
We ran each NER model on our created benchmark dataset, for all 139 years, and analyzed the performance of each template over female vs. male genders and compared the results of these models across genders per year. We report six sets of results based upon different concepts of error. The different errors we consider are discussed below. $N_f$ is the set of female names in a particular year. The same error is calculated for male using $N_m$ --- the set of male names.

\subsubsection{Error Type-1 Unweighted.} 
This is a type of error that measures names that are tagged as non-PERSON, or not tagged at all. In other words, any name not tagged as a PERSON is considered to be an error.
\[ \frac{\sum_{n \in N_f}I(n_{type} \neq PERSON)}{|N_f|}\]

\subsubsection{Error Type-1 Weighted.} This type of error is similar to Error Type-1 Unweighted; however, we considered how frequent the mistaken name is based on census data while calculating the error.
\[ \frac{\sum_{n \in N_f}freq_f(n_{type} \neq PERSON)}{\sum_{n \in N_f}freq_f(n)},\]
where $freq_f(\cdot)$ returns the frequency of a name in the female census data in a particular year. Similarly, $freq_m(\cdot)$ will yield the frequency of a name in the male census data.
Type-1 errors can be sub-divided into Type-2 and Type-3 errors and serve as a super-set for the following types.
\subsubsection{Error Type-2 Unweighted.} This is a type of error in which only names that are tagged,  but whose tags are non-PERSON are considered to be errors. This measure is similar to precision, where the model is only punished for incorrect retrieval. This error rate reports the percentage of names that are tagged as non-PERSON entities among all the names in a certain year.
\[ \frac{\sum_{n \in N_f}I(n_{type} \notin \{\emptyset,PERSON\})}{|N_f|},\]
where $\emptyset$ indicates the name is not tagged.
\subsubsection{Error Type-2 Weighted.} This type of error is similar to Error Type-2 Unweighted; however, here we considered how frequent the mistaken name was while calculating the error.
\[ \frac{\sum_{n \in N_f}freq_f(n_{type} \notin \{\emptyset,PERSON\})}{\sum_{n \in N_f}freq_f(n)}\]

\subsubsection{Error Type-3 Unweighted.} This is a type of error in which only names that are not tagged are considered to be errors. We do not consider names that are wrongfully tagged to non-PERSON as an error, but only names that are not tagged are considered erroneous. This error rate reports the percentage of names that are not tagged at all among all the names in a certain year.
\[ \frac{\sum_{n \in N_f}I(n_{type} = \emptyset)}{|N_f|}\]

\subsubsection{Error Type-3 Weighted.} 
\begin{figure*}[!bt]
\begin{subfigure}[b]{0.33\textwidth}
\includegraphics[width=\textwidth,height=0.65\textwidth,trim=0cm 0cm 0cm 0cm,clip=true]{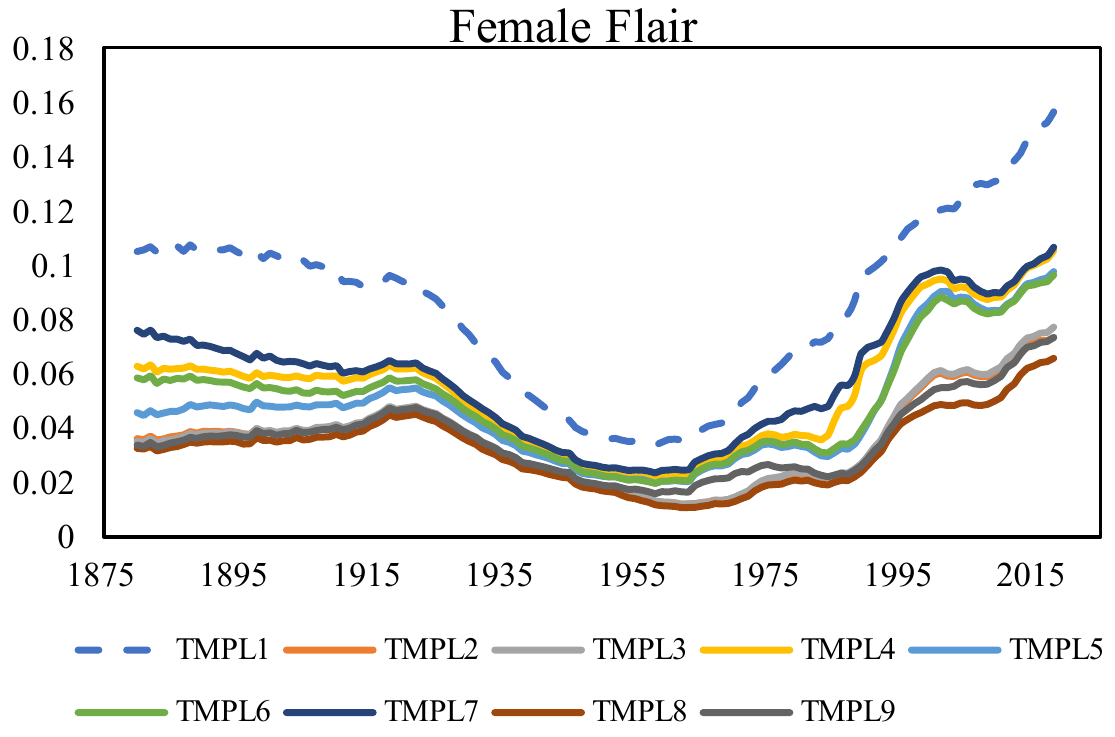}
\end{subfigure}
\begin{subfigure}[b]{0.33\textwidth}
\includegraphics[width=\textwidth,height=0.65\textwidth,trim=0cm 0cm 0cm 0cm,clip=true]{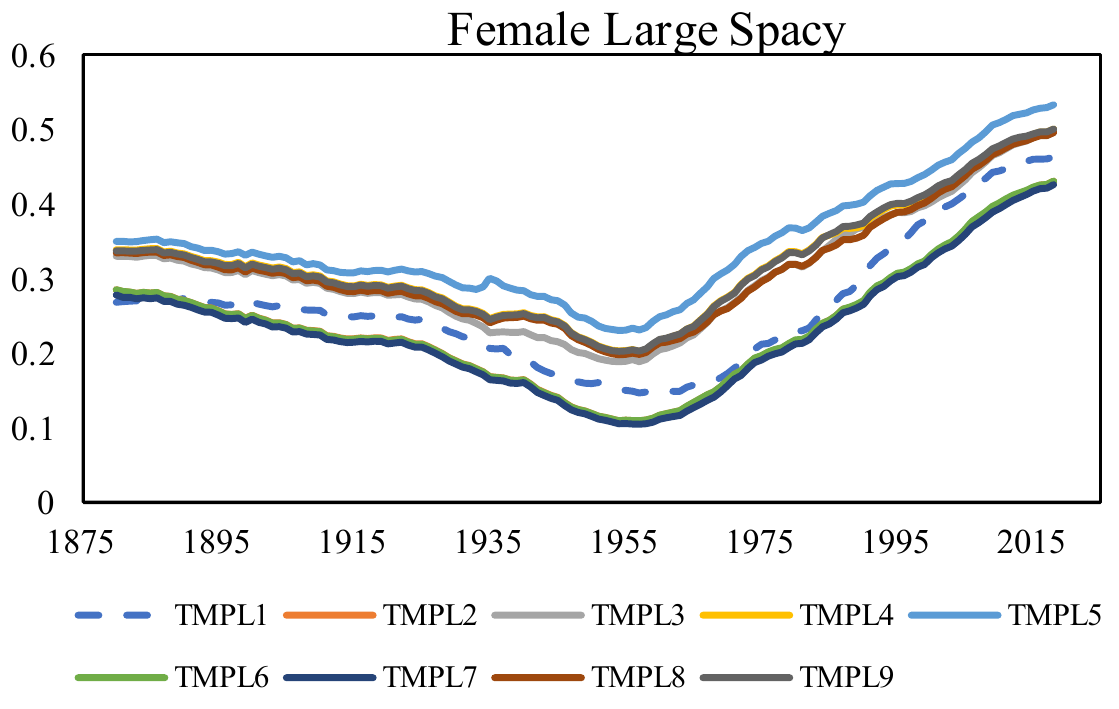}
\end{subfigure}
\begin{subfigure}[b]{0.33\textwidth}
\includegraphics[width=\textwidth,height=0.65\textwidth,trim=0cm 0cm 0cm 0cm,clip=true]{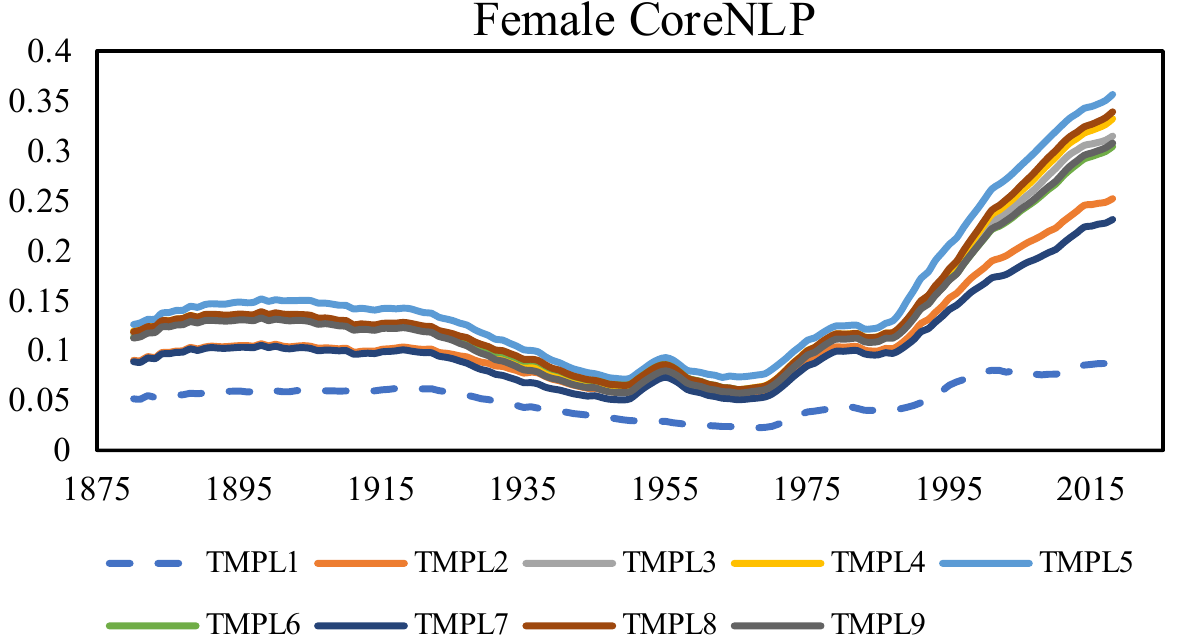}
\end{subfigure}
\begin{subfigure}[b]{0.33\textwidth}
\includegraphics[width=\textwidth,height=0.65\textwidth,trim=0cm 0cm 0cm 0cm,clip=true]{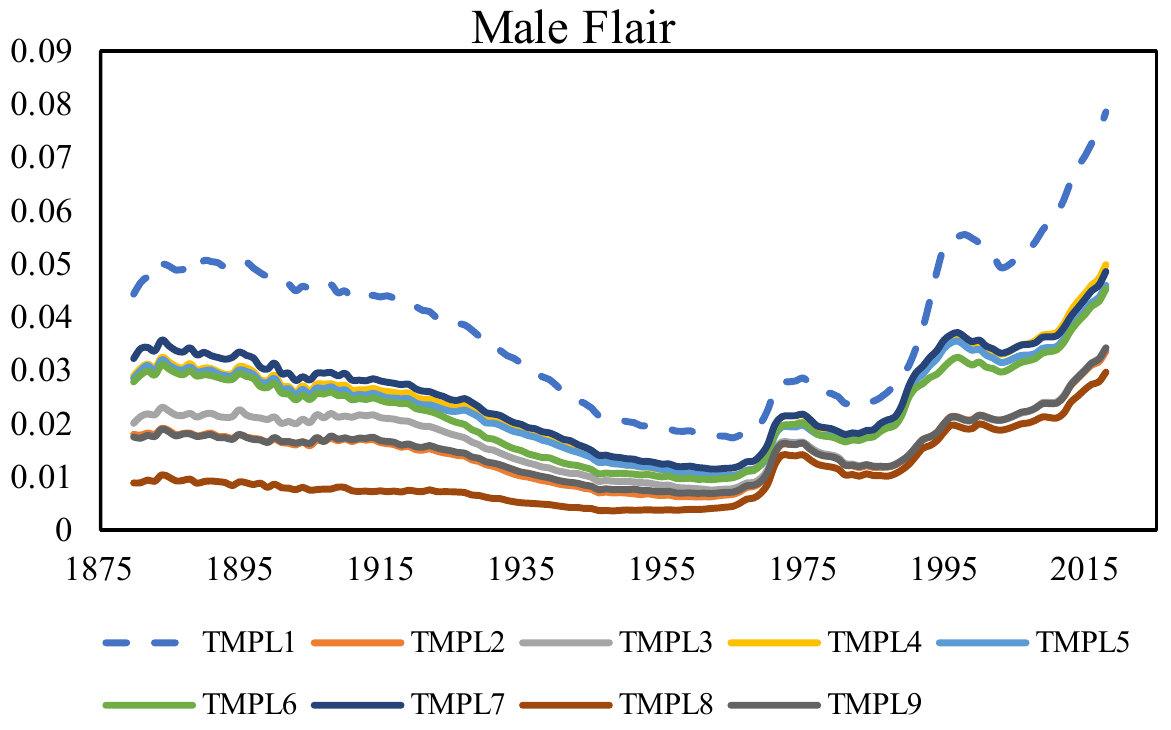}
\end{subfigure}
\begin{subfigure}[b]{0.33\textwidth}
\includegraphics[width=\textwidth,height=0.65\textwidth,trim=0cm 0cm 0cm 0cm,clip=true]{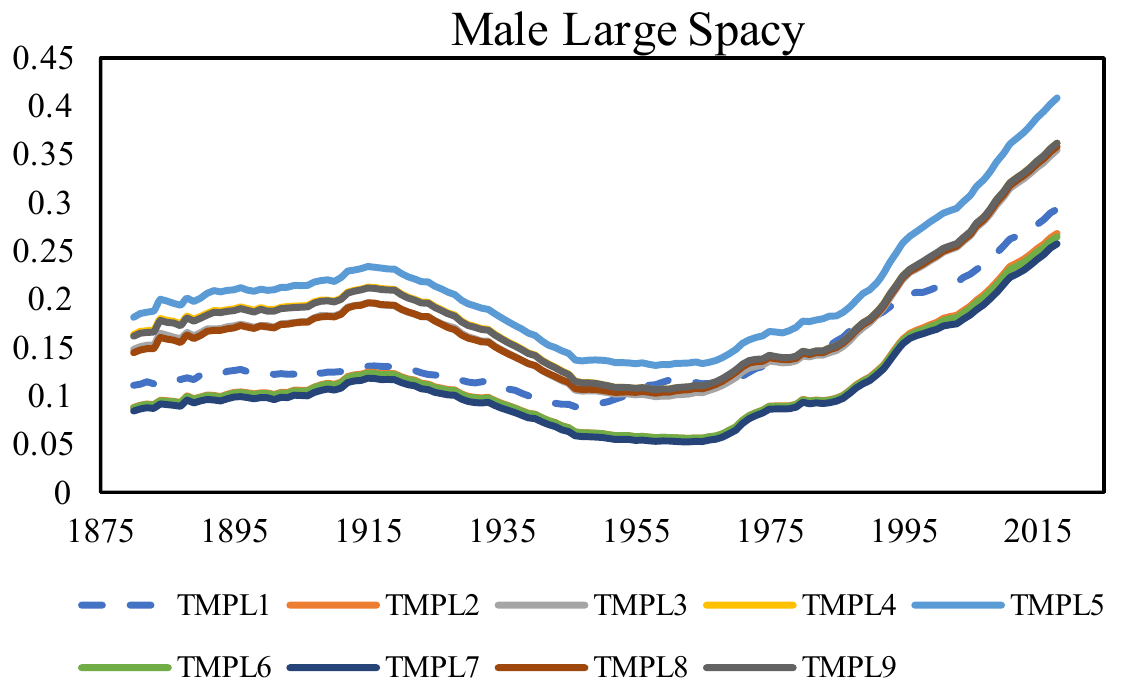}
\end{subfigure}
\begin{subfigure}[b]{0.33\textwidth}
\includegraphics[width=\textwidth,height=0.65\textwidth,trim=0cm 0cm 0cm 0cm,clip=true]{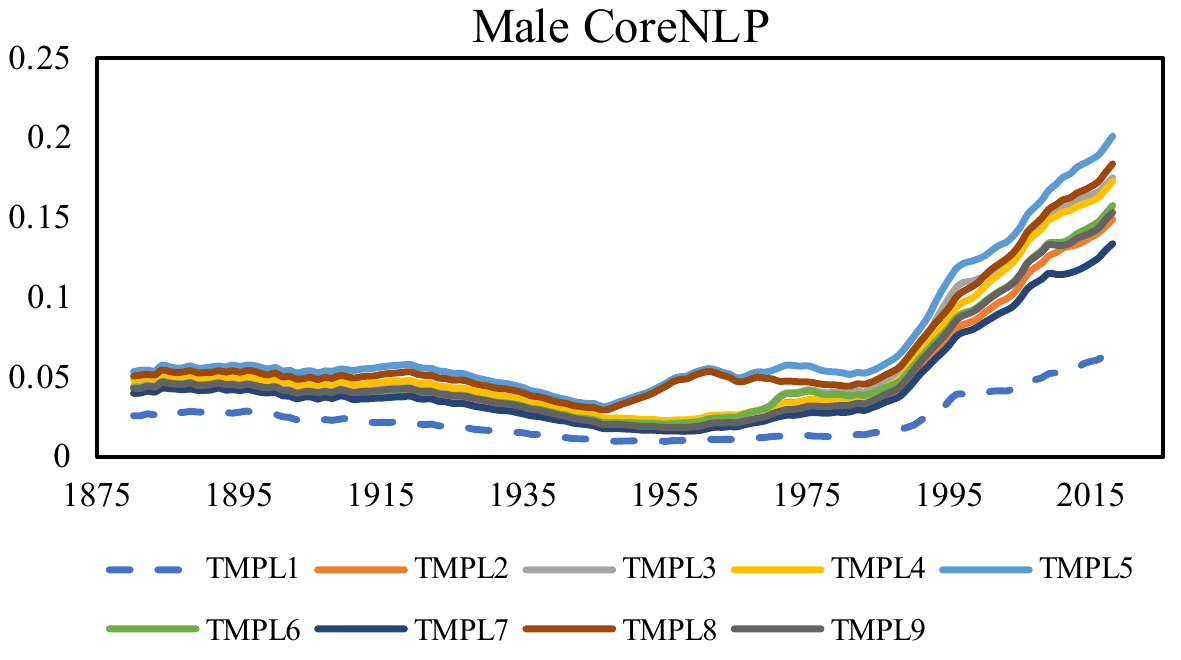}
\end{subfigure}
\caption{Performance of different models on different templates from our benchmark for female and male names collected for over 139 years. Notice how context in some of the templates helped some models, but showed negative effects on other models.}
\label{tempresults}
\end{figure*}

This type of error is similar to Error Type-3 Unweighted; however, we considered how frequent the mistaken name was while calculating the error.
\[ \frac{\sum_{n \in N_f}freq_f(n_{type} = \emptyset)}{\sum_{n \in N_f}freq_f(n)}\]
\begin{figure*}[!bt]
\begin{subfigure}[b]{0.33\textwidth}
\caption{Error Type-1 Unweighted}
\includegraphics[width=\textwidth,height=0.5\textwidth,trim=0cm 0cm 0cm 0cm,clip=true]{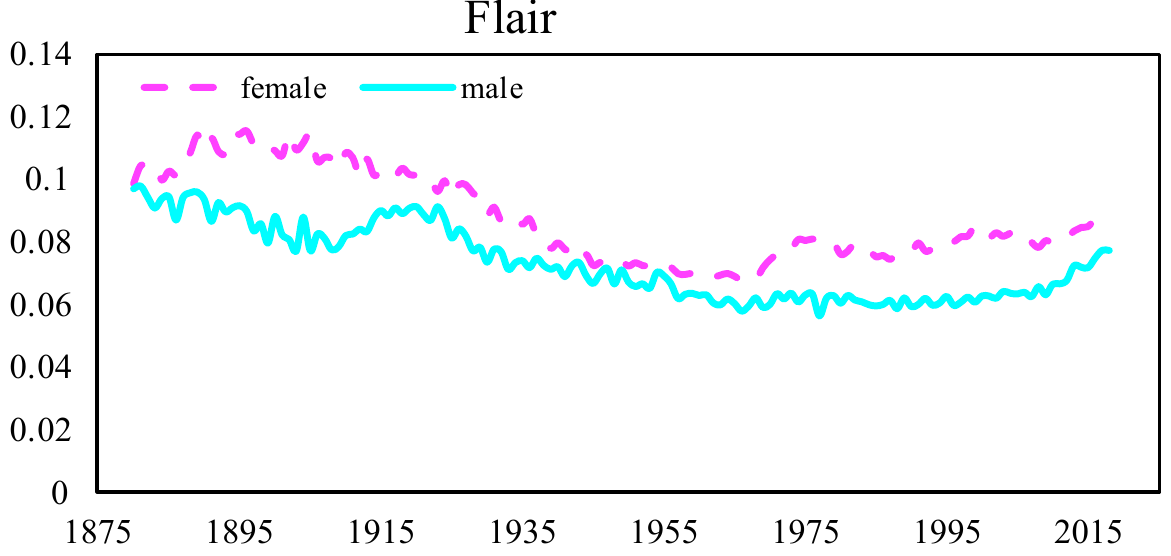}
\end{subfigure}
\begin{subfigure}[b]{0.33\textwidth}
\caption{Error Type-2 Unweighted}
\includegraphics[width=\textwidth,height=0.5\textwidth,trim=0cm 0cm 0cm 0cm,clip=true]{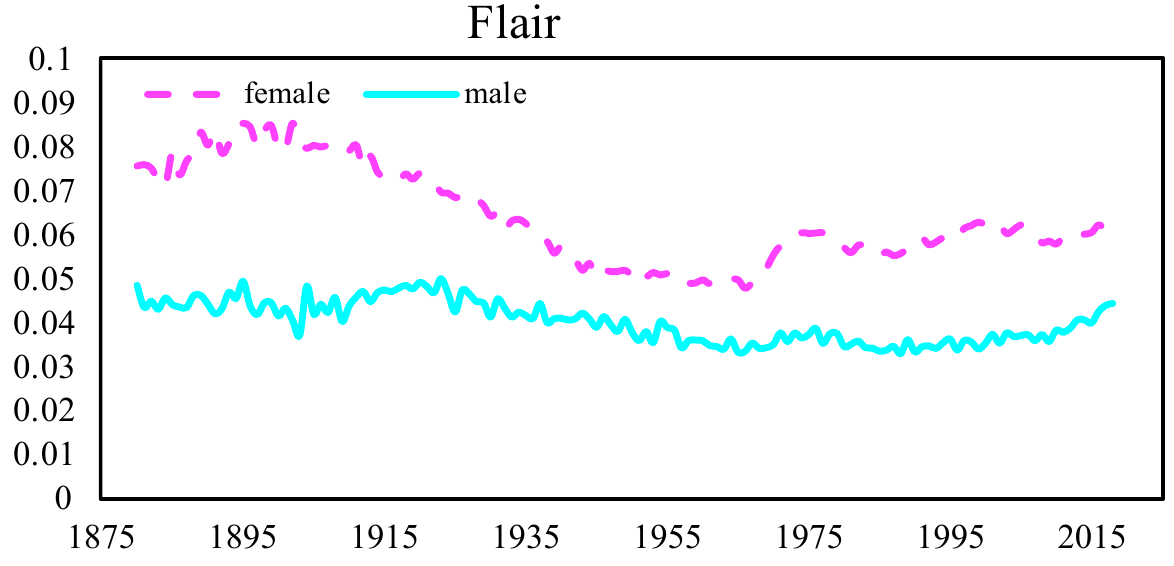}
\end{subfigure}
\begin{subfigure}[b]{0.33\textwidth}
\caption{Error Type-3 Unweighted}
\includegraphics[width=\textwidth,height=0.5\textwidth,trim=0cm 0cm 0cm 0cm,clip=true]{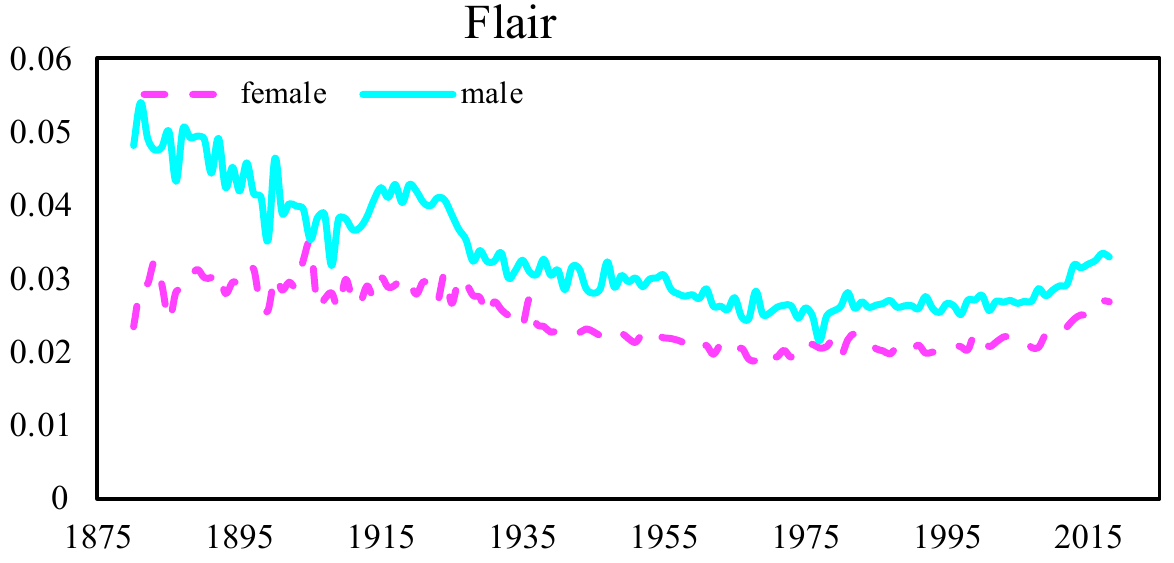}
\end{subfigure}
\begin{subfigure}[b]{0.33\textwidth}
\includegraphics[width=\textwidth,height=0.5\textwidth,trim=0cm 0cm 0cm 0cm,clip=true]{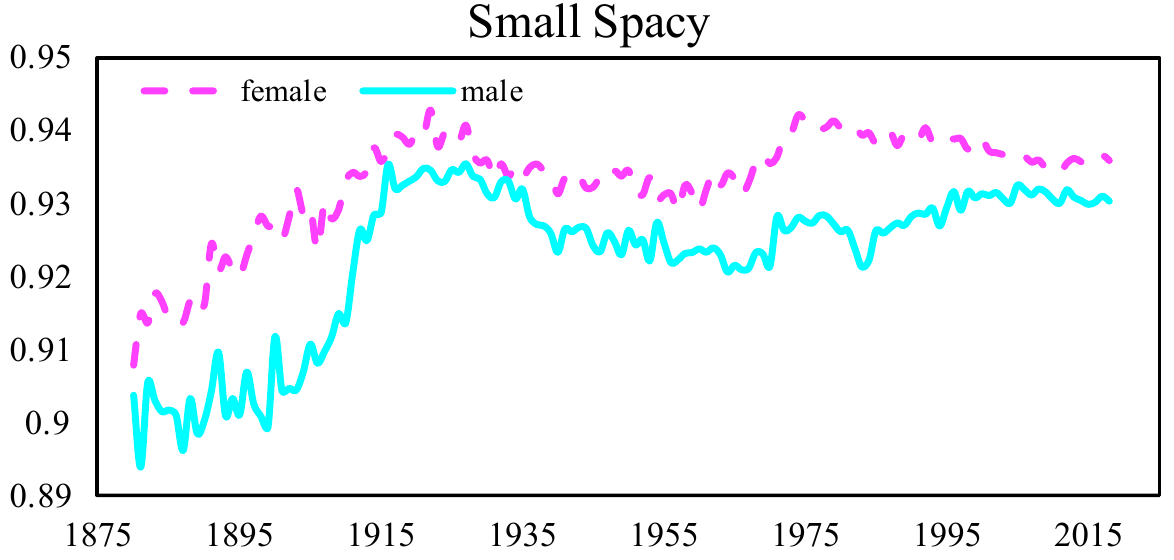}
\end{subfigure}
\begin{subfigure}[b]{0.33\textwidth}
\includegraphics[width=\textwidth,height=0.5\textwidth,trim=0cm 0cm 0cm 0cm,clip=true]{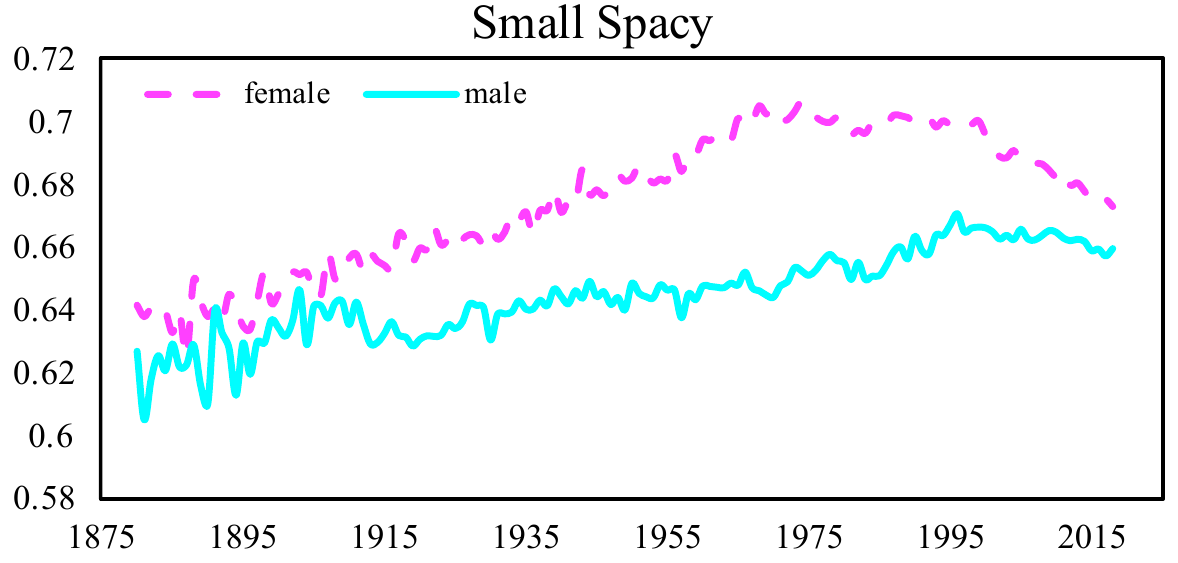}
\end{subfigure}
\begin{subfigure}[b]{0.33\textwidth}
\includegraphics[width=\textwidth,height=0.5\textwidth,trim=0cm 0cm 0cm 0cm,clip=true]{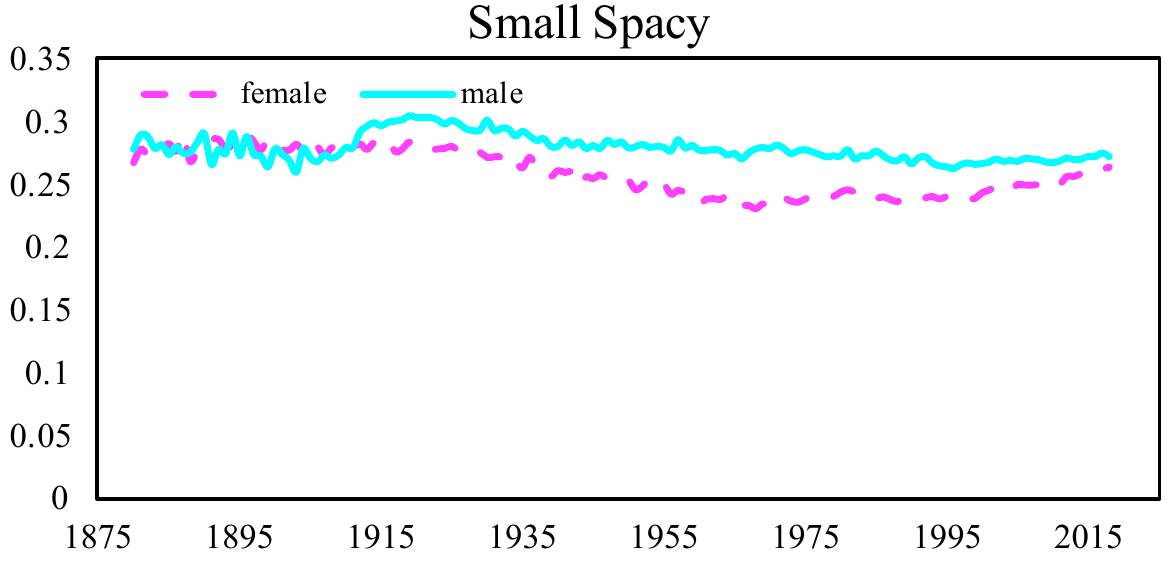}
\end{subfigure}
\begin{subfigure}[b]{0.33\textwidth}
\includegraphics[width=\textwidth,height=0.5\textwidth,trim=0cm 0cm 0cm 0cm,clip=true]{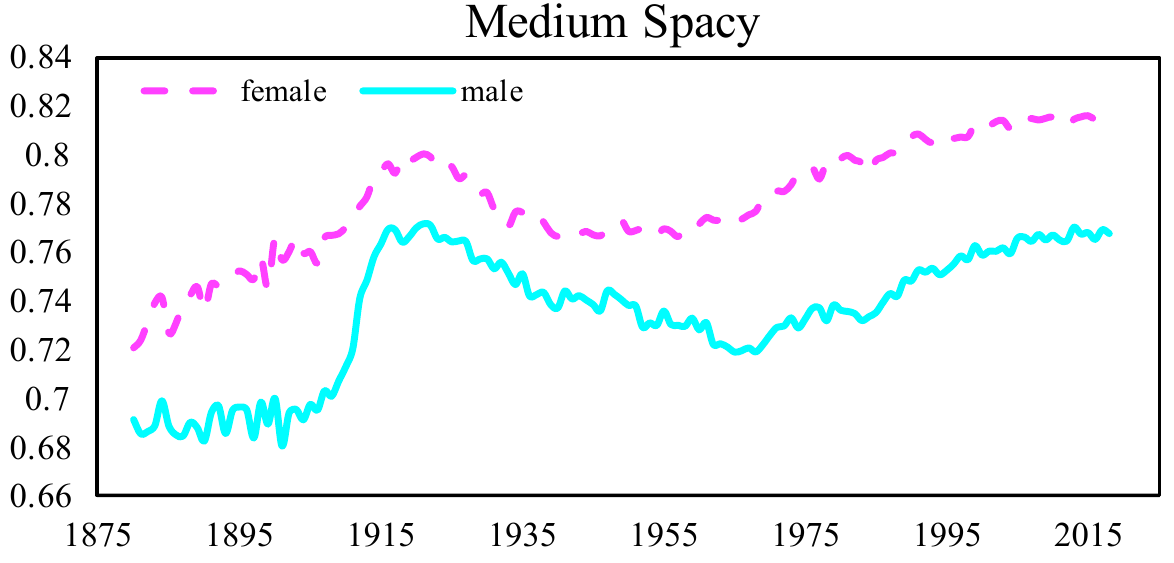}
\end{subfigure}
\begin{subfigure}[b]{0.33\textwidth}
\includegraphics[width=\textwidth,height=0.5\textwidth,trim=0cm 0cm 0cm 0cm,clip=true]{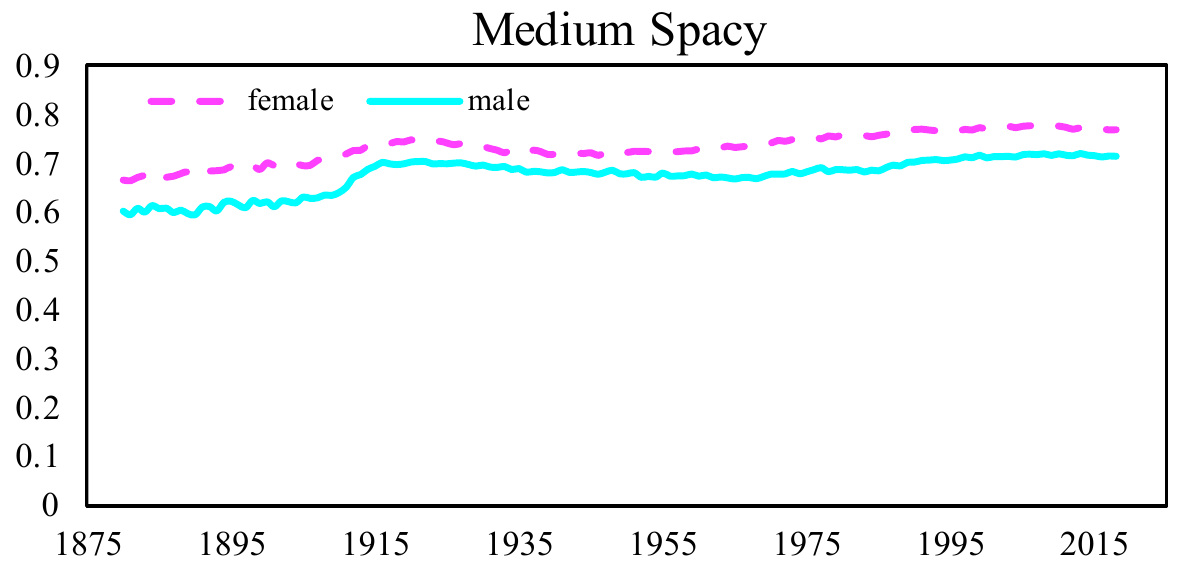}
\end{subfigure}
\begin{subfigure}[b]{0.33\textwidth}
\includegraphics[width=\textwidth,height=0.5\textwidth,trim=0cm 0cm 0cm 0cm,clip=true]{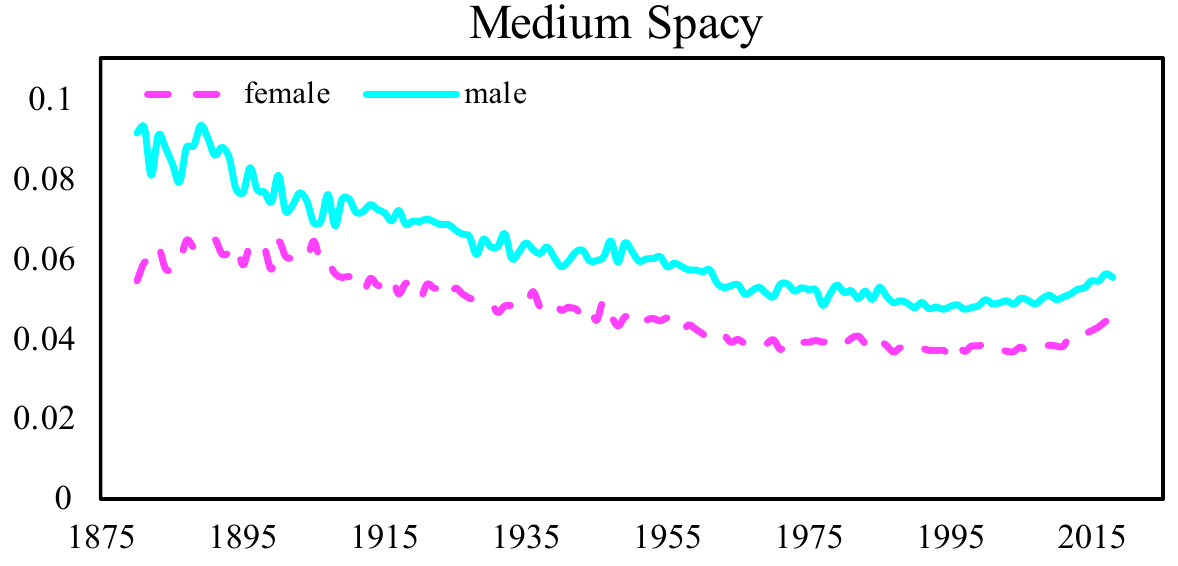}
\end{subfigure}
\begin{subfigure}[b]{0.33\textwidth}
\includegraphics[width=\textwidth,height=0.5\textwidth,trim=0cm 0cm 0cm 0cm,clip=true]{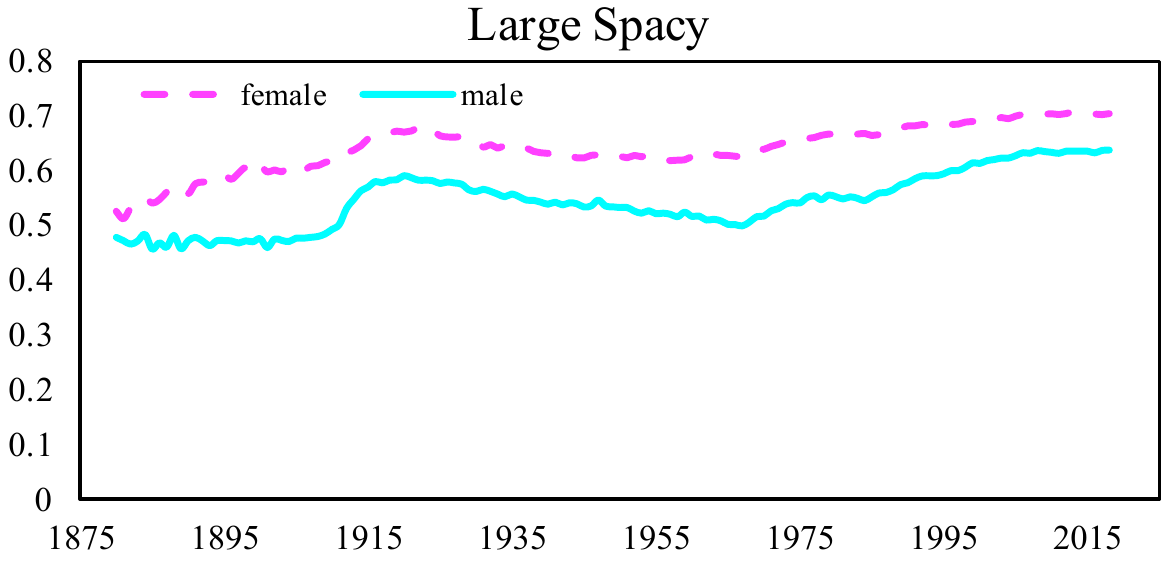}
\end{subfigure}
\begin{subfigure}[b]{0.33\textwidth}
\includegraphics[width=\textwidth,height=0.5\textwidth,trim=0cm 0cm 0cm 0cm,clip=true]{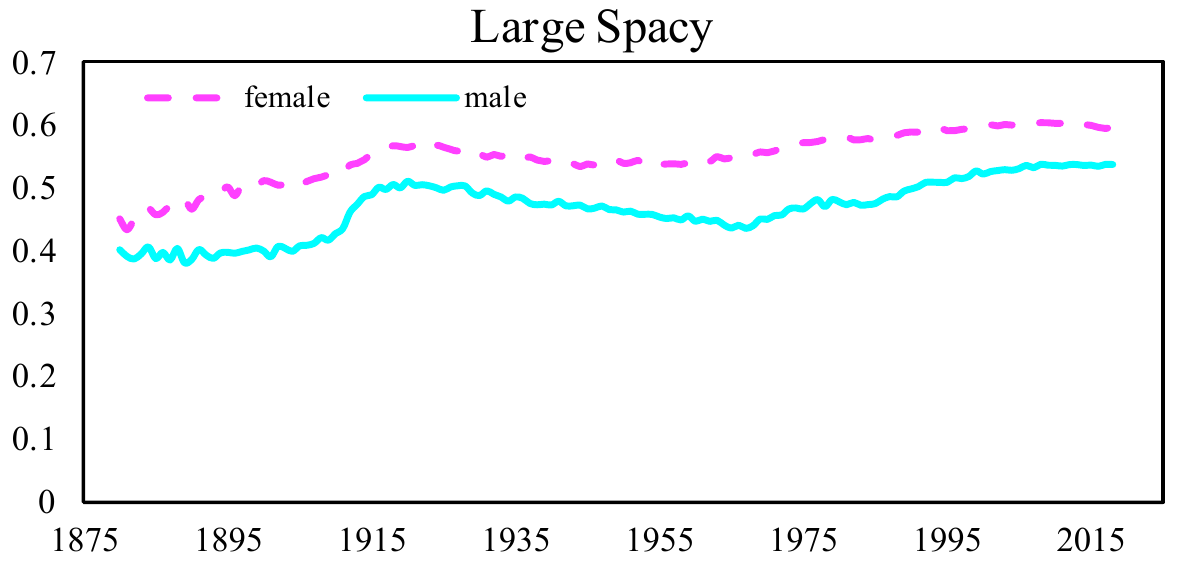}
\end{subfigure}
\begin{subfigure}[b]{0.33\textwidth}
\includegraphics[width=\textwidth,height=0.5\textwidth,trim=0cm 0cm 0cm 0cm,clip=true]{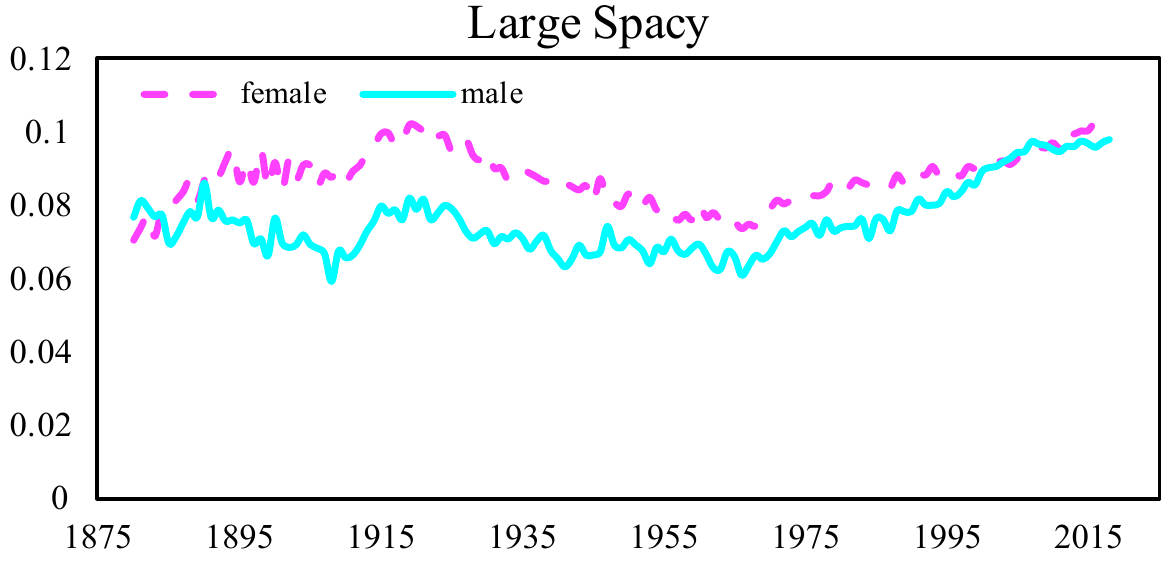}
\end{subfigure}
\begin{subfigure}[b]{0.33\textwidth}
\includegraphics[width=\textwidth,height=0.5\textwidth,trim=0cm 0cm 0cm 0cm,clip=true]{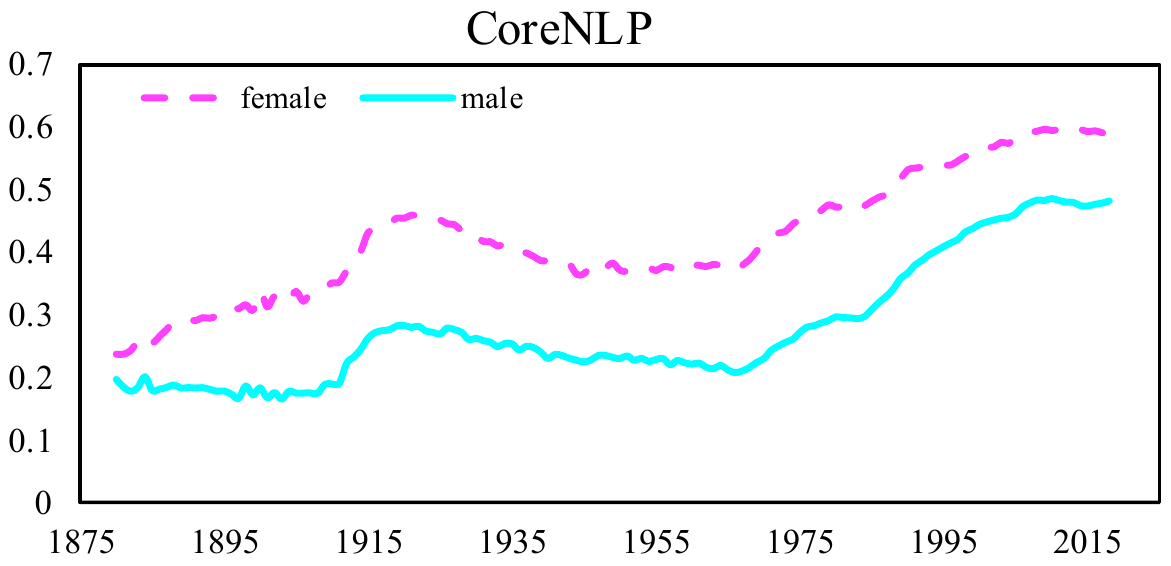}
\end{subfigure}
\begin{subfigure}[b]{0.33\textwidth}
\includegraphics[width=\textwidth,height=0.5\textwidth,trim=0cm 0cm 0cm 0cm,clip=true]{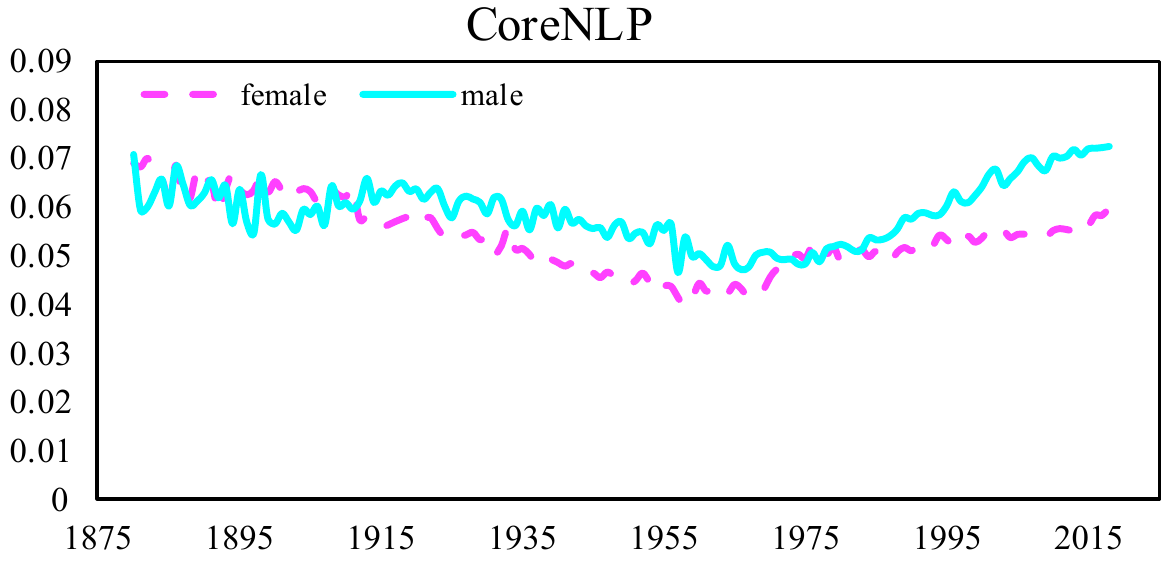}
\end{subfigure}
\begin{subfigure}[b]{0.33\textwidth}
\includegraphics[width=\textwidth,height=0.5\textwidth,trim=0cm 0cm 0cm 0cm,clip=true]{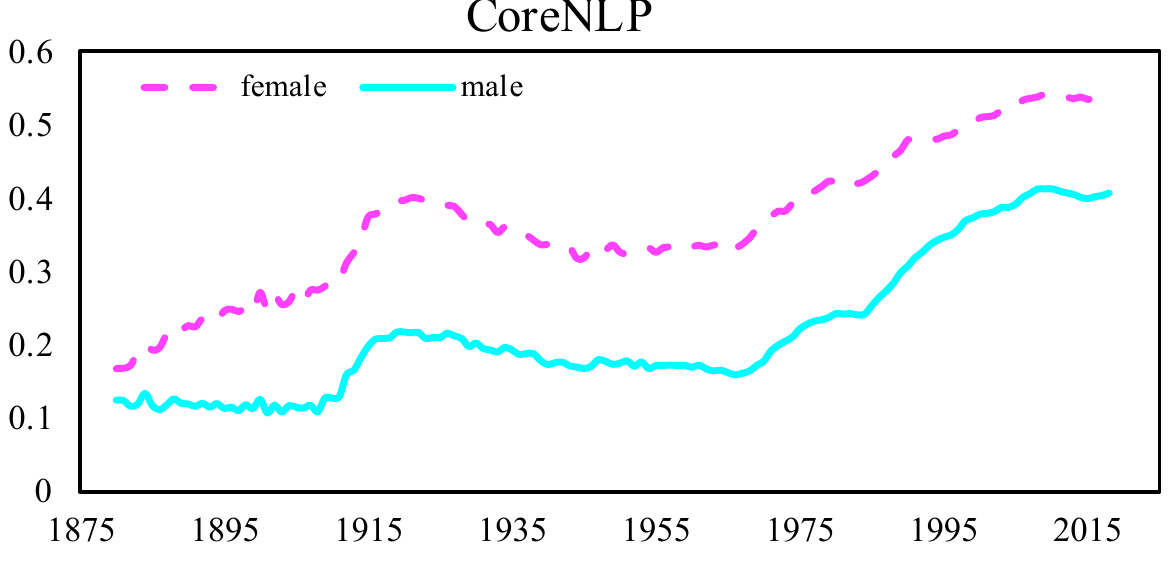}
\end{subfigure}
\caption{Results from different models over the  139-year history of baby names from the census data on different error types for the unweighted cases using template \#4. Female names have higher error rates for all cases except four marginal cases. The y axis shows the calculated error rates for each of the error types as described in their corresponding formulas, and the x axis represents the year in which the baby name was given.}
\label{result2}
\end{figure*}
Different types of errors allow for fine-grained analysis into the existence of different biases. Our results indicate that all models are mostly more biased toward female names vs. male names, as shown in Figures \ref{result1} and \ref{result2} over the 139-year history. The fact that all the weighted cases are biased toward female names shows that more frequent and popular female names are susceptible to bias and error in named entity recognition systems---which is a more serious type of error to consider. For space considerations, we only report the results for one of the templates (Template \#4) since the results were following similar trend for all the other templates wherein the models were mostly more biased toward female names. We have included results from other templates in our supplementary material to demonstrate that other templates also follow a similar pattern. That being said, in Figure \ref{tempresults} we showed how all the models perform on all the templates, for Error Type-1 Weighted (the super-set error) case, which on its own is showing an interesting trend. This result shows how context helps some models over others by bringing down the error rates when sentences are added to the names (templates \#2 through \#9). Other models perform better on template \#1, showing that context, in fact confuses the model. We observe that in contextual-based models such as Flair, context indeed helps the model make the right decisions by it having less error rates for templates \#2 through \#9 compared to template \#1. Other models do not necessarily follow this pattern. 
As an example, we provide the types of names and errors that can happen in these models. We list the top six most frequent male and female names which were tagged erroneously by the Flair model in year 2018 from our benchmark evaluated on template \#4 in Table \ref{examples}.
\section{Model Version Evaluation and Comparison}
Updates to models often lead to superior performance; however, this may come with a detriment to fairness and bias. In this section we want to see how much the version updates in models will be robust toward fairness constraints and errors defined in the previous section. Thus, we will first define this source of bias. Then we will show the results for the models from the previous section that have various versions.  
\\
\textbf{Definition} \textit{(Version Bias). This is a type of bias that  arises from updates in the systems.}

\begin{figure*}[!bt]
\begin{subfigure}[b]{0.33\textwidth}
\caption{Error Type-2 Weighted}
\includegraphics[width=\textwidth,height=0.55\textwidth,trim=0cm 0cm 0cm 0cm,clip=true]{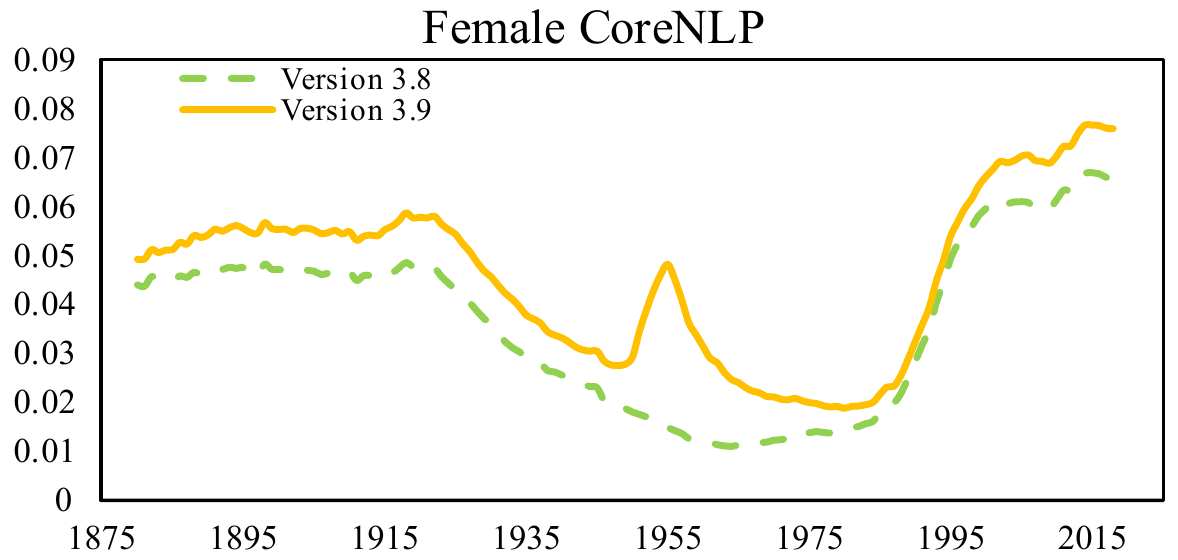}
\end{subfigure}
\begin{subfigure}[b]{0.33\textwidth}
\caption{Error Type-3 Weighted}
\includegraphics[width=\textwidth,height=0.55\textwidth,trim=0cm 0cm 0cm 0cm,clip=true]{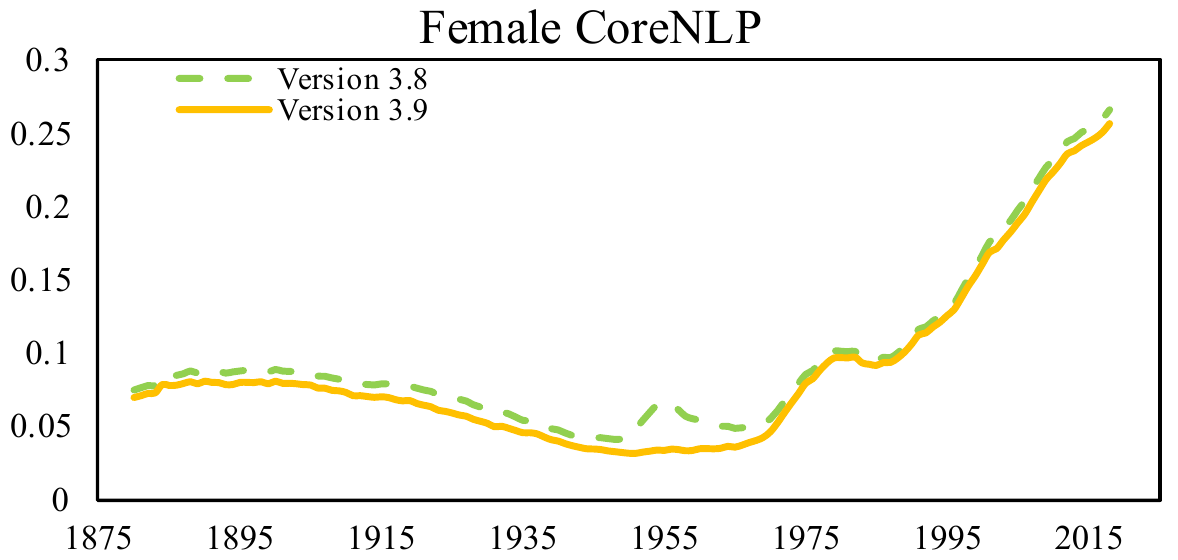}
\end{subfigure}
\begin{subfigure}[b]{0.33\textwidth}
\caption{Error Type-1 Weighted}
\includegraphics[width=\textwidth,height=0.55\textwidth,trim=0cm 0cm 0cm 0cm,clip=true]{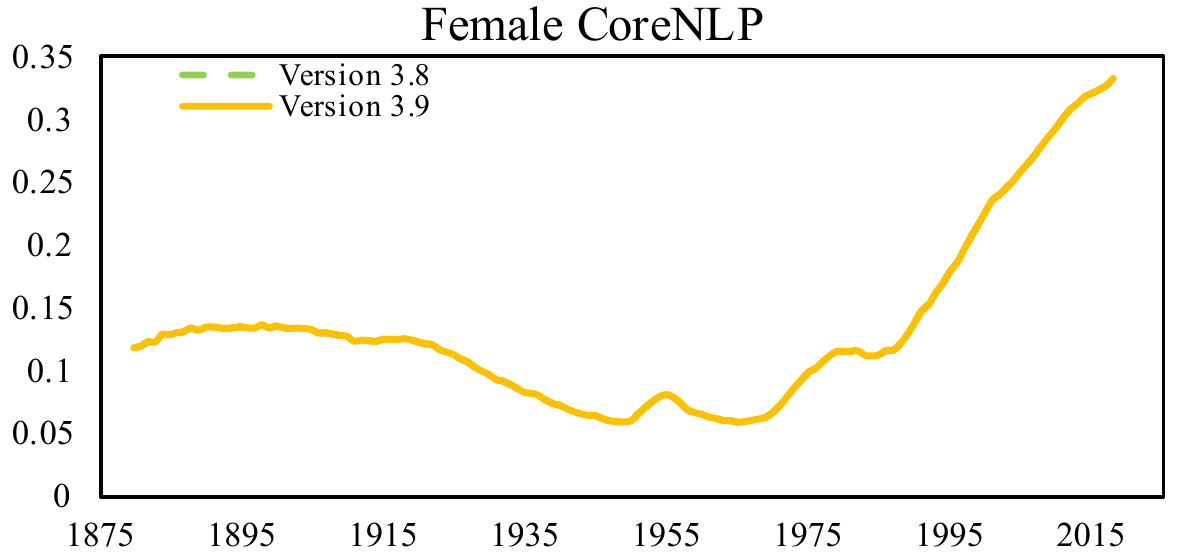}
\end{subfigure}
\begin{subfigure}[b]{0.33\textwidth}
\includegraphics[width=\textwidth,height=0.55\textwidth,trim=0cm 0cm 0cm 0cm,clip=true]{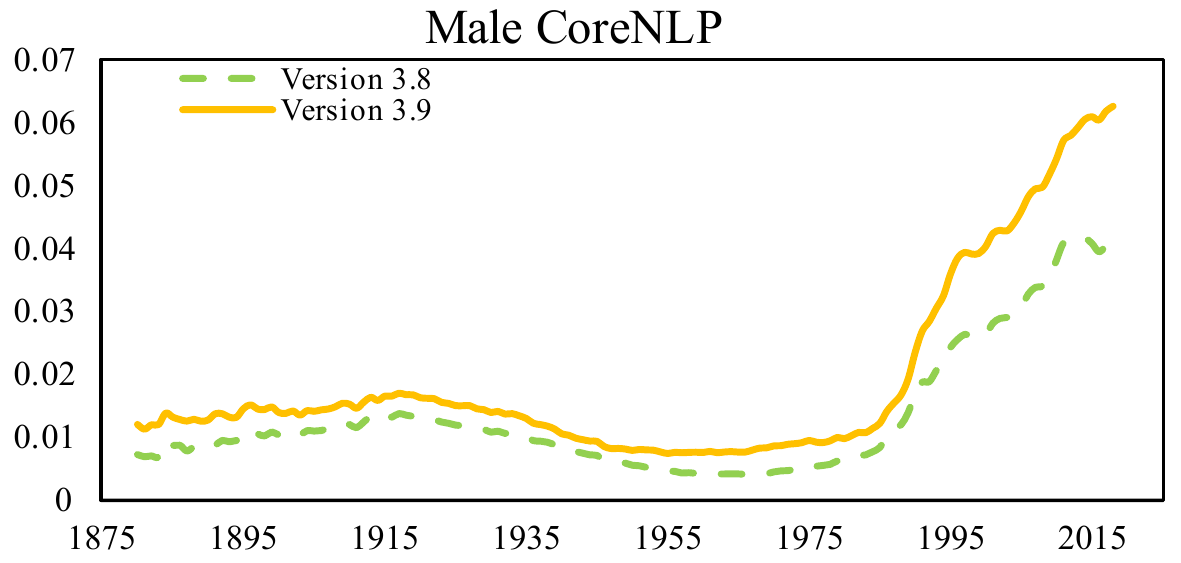}
\end{subfigure}
\begin{subfigure}[b]{0.33\textwidth}
\includegraphics[width=\textwidth,height=0.55\textwidth,trim=0cm 0cm 0cm 0cm,clip=true]{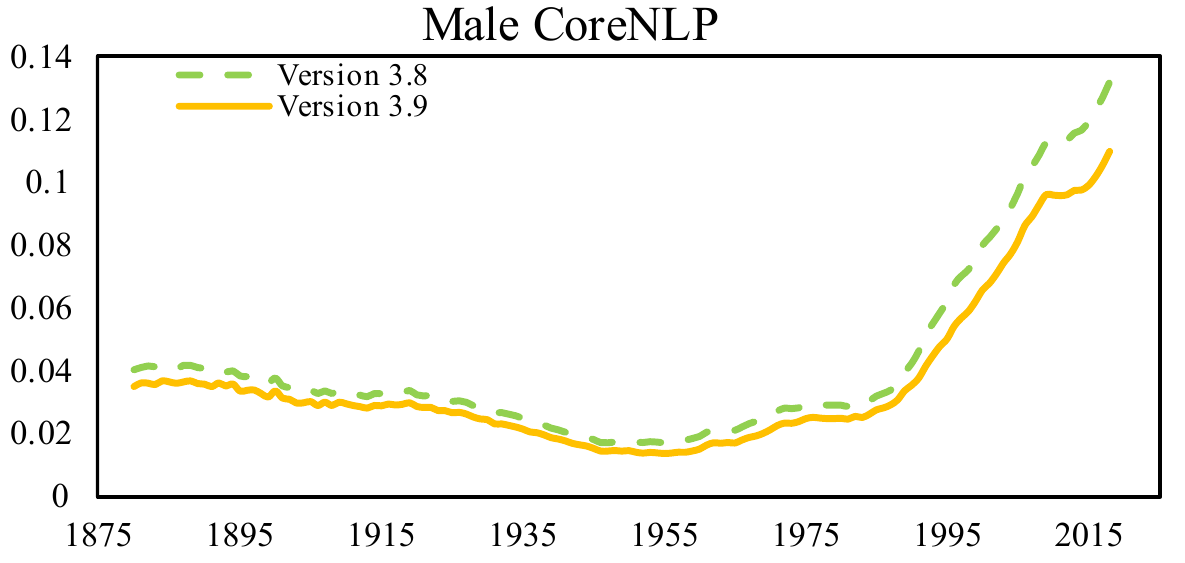}
\end{subfigure}
\begin{subfigure}[b]{0.33\textwidth}
\includegraphics[width=\textwidth,height=0.55\textwidth,trim=0cm 0cm 0cm 0cm,clip=true]{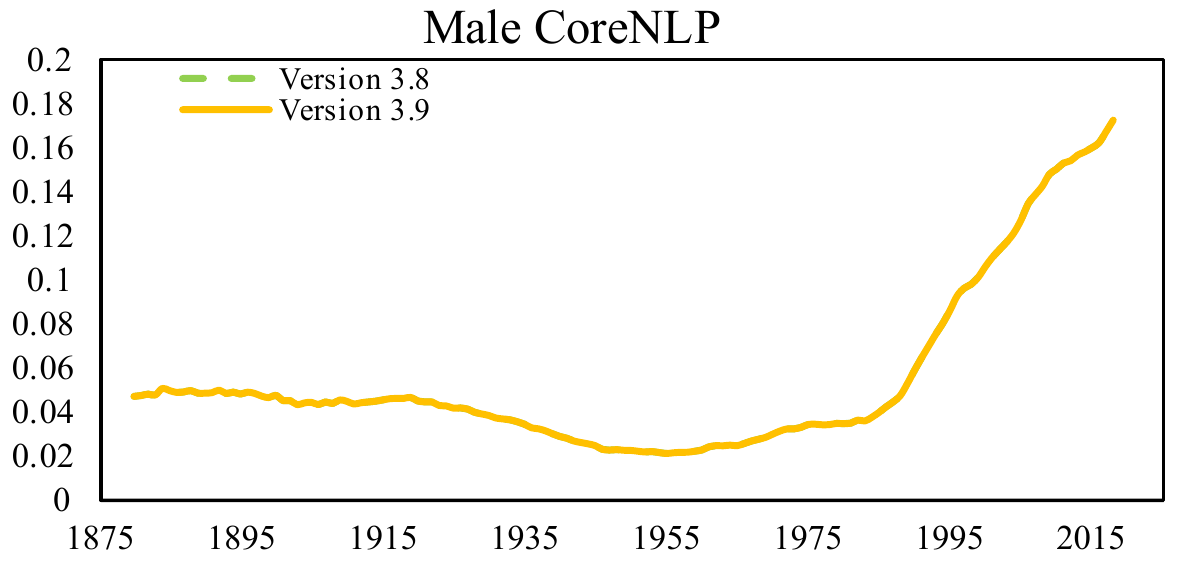}
\end{subfigure}
\caption{Version update in the CoreNLP model tried to tag more entities and thus a subtle boost in performance with regard to Error Type-3. However, this resulted in more PERSON entities being tagged as non-PERSON entities.  This then degraded performance with regard to Error Type-2 in the newer version, and overall no change in the Error Type-1 case which is considered to be the super-set. We observe how the degrade in Error Type-2 affected more female names than males on average.}
\label{corenlp_versions}
\end{figure*}

\begin{figure}[t]
\includegraphics[width=0.45\textwidth,height=0.2\textwidth,trim=0cm 0cm 0cm 0cm,clip=true]{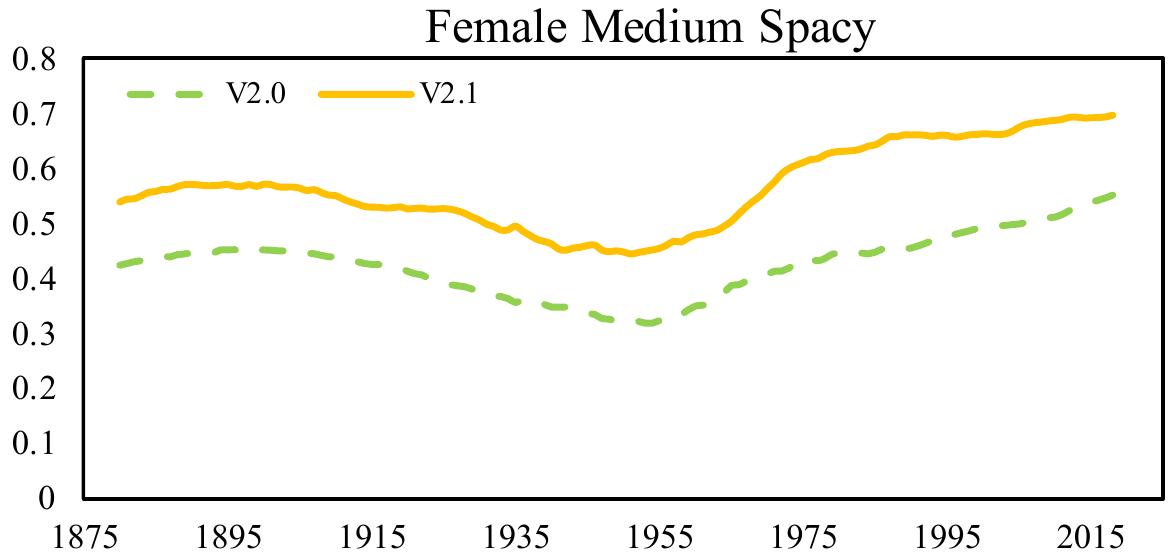}
\includegraphics[width=0.45\textwidth,height=0.2\textwidth,trim=0cm 0cm 0cm 0cm,clip=true]{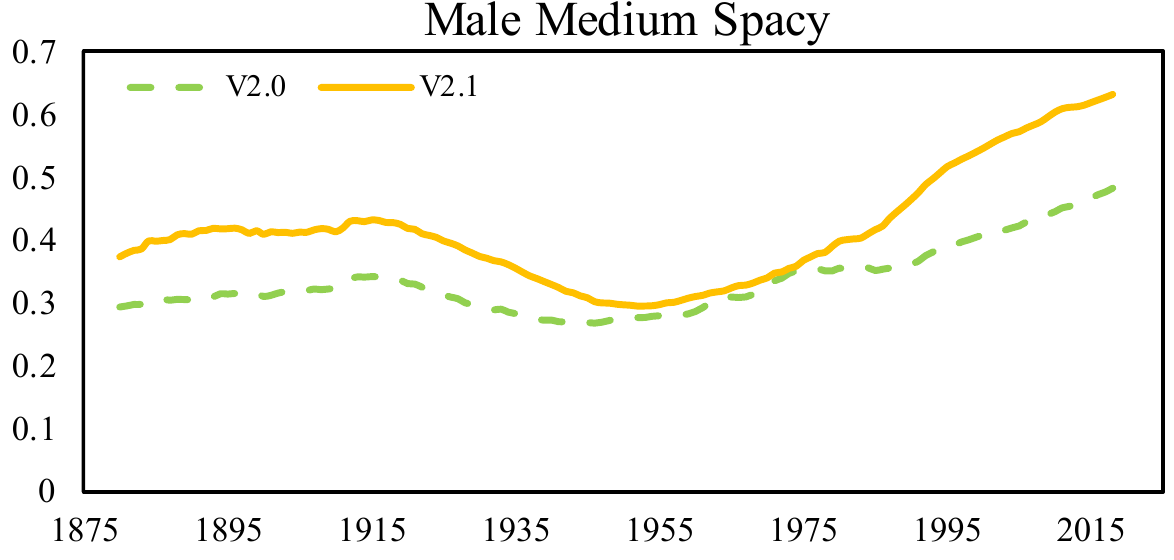}
\caption{Biased performance of version 2.1 over 2.0 in Spacy medium. The bias against female names was on average twice that of male names.}
\label{spacyv}
\end{figure}

In order to report the results of our analysis, we used four of our models mentioned in the previous section that have version updates---namely small, medium, and large models from Spacy (versions 2.0 and 2.1) and versions 3.8 and 3.9 from CoreNLP. We then repeated the experiments from the previous section to report the results for Error Type-1, 2, and 3 (Weighted and Unweighted) cases using template \#4 from our benchmark. The results for Spacy show that although in not all cases were updated versions worse than that of the previous versions, there were some cases where  version updates had serious fairness-related issues. For instance, as shown in Figure \ref{spacyv} for the Spacy medium model, the newer version is more erroneous when it comes to the Error Type-1 Weighted case which is a superset of all the error types discussed in this paper. Not only that, the average increase rate of this error is twice that of female names as opposed to male names when updating the model version. These newer versions may try to boost their performance by trying to tag more entities, which indeed would boost their performance when having other non-PERSON entities in the test set. However, these newer tags may be tagging PERSON entities in the relevant context to the non-PERSON counterpart, which is not desirable when evaluated on our benchmark. The reason that we separated the types of errors and created the benchmark is to show the fine-grained and sensitive issues in these systems. Similarly, for CoreNLP more entities tried to be tagged---which resulted in a slight improvement in Error Type-3, as shown in Figure \ref{corenlp_versions}. However, these tags would not correctly assign PERSON tags to PERSON entities, which resulted in a  degraded performance in Error Type-2, as shown in Figure \ref{corenlp_versions} and 
resulted in no change in the overall performance based on Error Type-1 in the newer version update of the CoreNLP model. The results were identical for the unweighted case, and we can see how that degrade in performance was more severe for female names on average. As examples, some of the changes from version 3.8 to version 3.9 of the CoreNLP model are shown in Table \ref{corenlp_version_examples}.

\section{Bias in Data}
Since data plays an important role in the outcome of the model and can directly affect the fairness constraints if it contains any biases, we decided to analyze some of the datasets that are widely used in the training of NER models to determine whether they show any biases toward a specific group that could result in the biased behavior observed in those results discussed in previous sections. 
\begin{table}
\begin{tabular}{ p{2.6cm}p{1.4cm}p{2.4cm}}
 \toprule
 Female Name&Frequency&Error Type\\
 \midrule
 Charlotte&12,940&Tagged as LOC\\[0.5pt]
 Sofia&7,621&Tagged as LOC\\[0.5pt]
 Victoria&7,089&Tagged as LOC\\[0.5pt]
 Madison&7,036&Tagged as LOC\\[0.5pt]
 Aurora&4,785&Tagged as LOC\\[0.5pt]
 \bottomrule
\end{tabular}
\newline
\vspace*{0.3 cm}
\newline
\begin{tabular}{ p{2.6cm}p{1.4cm}p{2.4cm}}
 \toprule
 Male Name&Frequency&Error Type\\
 \midrule
 Christian&6,509&Tagged as MISC\\[0.5pt]
 Jordan&4,646&Tagged as LOC\\[0.5pt]
 Roman&4,364&Tagged as MISC\\[0.5pt]
 Kaiden&2,832&Tagged as LOC\\[0.5pt]
 King&2,579&Not Tagged\\[0.5pt]
 \bottomrule
\end{tabular}
\caption{Top 5 mistagged examples from the Flair model on Template \#4 of female and male names from our benchmark.}

\label{examples}
\end{table}
We used the train, test, and development sets from two widely known CoNLL-2003\footnote{\small{\url{https://www.clips.uantwerpen.be/conll2003/ner/}}} \cite{sang2003introduction} and OntoNotes-5\footnote{\small{\url{https://catalog.ldc.upenn.edu/LDC2013T19}}} \cite{weischedel2012ontonotes} datasets which were used in the training and testing of Flair, Spacy, and many other models. The split of the OntoNotes-5 dataset into train, development, and test sets was performed according to \cite{pradhan2013towards}. We reported the percentages of male vs. female names from the census data that appeared in train, test, and development sets in each of the datasets and compared this to the percentages of male vs. female names in reality from the census data to see how much these datasets are reflective of the reality or if they pertain to any bias toward a specific gender group.

Our results shown in Table \ref{data_stats_table} indicate that the datasets used do not reflect the real world, but rather exactly the opposite of that. Unlike the census data, which is representative of real-world statistics, wherein female names have more versatility---62\% unique names vs. 38\% unique male names---datasets used in training the NER models contain 42\% female names vs. 58\% male names from the census data. Not only do the datasets not contain more versatile female names to reflect the reality,  but instead have even less variety which can itself bias the models by not covering enough female names. Similar patterns are observable in test and development sets of datasets used in the NER systems.

\section{Related Work}
We have seen large amounts of attention and work regarding fairness in machine learning and natural language processing models and methods. Recent papers observed a type of stereotyping bias in word embedding methods and tried to mitigate this type of bias by proposing a method that respects the embeddings for gender-specific words but de-biases embeddings for gender-neutral words \cite{bolukbasi2016man}, or by generating a gender-neutral version of Glove (called GN-Glove) that aims to preserve gender information in some directions of word vectors, while setting other dimensions free from gender influence \cite{zhao2018learning} or other data augmentation techniques \cite{pmlr-v97-brunet19a,zhao2019gender}. Other work tried to show and address bias in co-reference resolution \cite{zhao2018gender}, semantic role labeling \cite{zhao2017men}, machine translation \cite{font2019equalizing}, language models \cite{bordia2019identifying}, and sentence embedding \cite{may2019measuring}.

Addressing fairness and bias, not only in NLP but also in general machine learning, has lately gained much attention. 
\begin{table}
\begin{tabular}{ p{1.2cm} p{3.3cm} p{3.2cm}}
 \toprule
 Female Name&CoreNLP version 3.8&CoreNLP version 3.9\\
 \midrule
 Isabel&Not Tagged&Tagged as MISC\\[0.5pt]
 Angelina&Not Tagged&Tagged as MISC\\[0.5pt]
 June&Not Tagged&Tagged as DATE\\[0.5pt]
 Charlotte&Tagged as LOCATION&Tagged as CITY\\[0.5pt]
 Victoria&Tagged as LOCATION&Tagged as CITY\\[0.5pt]
 Sydney&Tagged as LOCATION&Tagged as CITY\\[0.5pt]
  \bottomrule
\end{tabular}
\newline
\vspace*{0.3 cm}
\newline
\begin{tabular}{ p{1.2cm} p{3.3cm} p{3.2cm}}
 \toprule
 Male Name&CoreNLP version 3.8&CoreNLP version 3.9\\
 \midrule
 Christian&Not Tagged&Tagged as RELIGION\\[0.5pt]
 Logan&Tagged as LOCATION&Tagged as CITY\\[0.5pt]
 Roman&Not Tagged&Tagged as MISC\\[0.5pt]
 Santiago&Tagged as LOCATION&Tagged as CITY\\[0.5pt]
 Jordan&Tagged as LOCATION&Tagged as CITY\\[0.5pt]
 Messiah&Not Tagged&Tagged as TITLE\\[0.5pt]
 \bottomrule
\end{tabular}
\caption{Some examples on how tagging changed during version update of the CoreNLP model. Note how the original problem of tagging PERSON entities correctly has not been addressed.}
    \label{corenlp_version_examples}
\end{table}
In \citet{mehrabi2019survey}, the authors created a taxonomy on fairness and bias that discusses how researchers have addressed fairness related issues in different fields. From representation learning \cite{moyer2018invariant} to graph embedding \cite{bose2019compositional} to community detection \cite{mehrabi2019debiasing} and clustering \cite{pmlr-v97-backurs19a}, researchers have studied biases in these areas and tried to address them by pointing out the observed problems and proposing new directions and ideas. In \citet{buolamwini2018gender} authors show and analyze the existing gender bias in facial recognition systems, such as those used by IBM, Microsoft, and Face++, and created a benchmark for better evaluation of bias in facial recognition systems. This is considered a significant contribution as it opens many future research questions and related papers. Paying attention to different AI applications and pointing out their issues in terms of fairness is an important issue that needs serious attention for significant future improvements to these systems.





\section{Conclusion and Future Work}
\begin{table}
\begin{tabular}{ p{0.3cm} p{2cm} p{1.0cm} p{1.0cm} p{1cm} p{1cm}}
 \toprule
& \textbf{Dataset} & \textbf{Female Count} &\textbf{Male Count}&\textbf{Female Pct}& \textbf{Male Pct}\\
 \midrule
 & Census & 67,698 & 41,475 & 62\% & 38\% \\[0.5pt]
 \midrule
 \parbox[t]{2mm}{\multirow{2}{*}{\rotatebox[origin=c]{90}{Train}}}
 & CoNLL 2003& 1,810 & 2,506&42\%&58\%\\[0.5pt]
 & OntoNotes5& 2,758&3,832&42\%&58\%\\[0.5pt]
 \midrule
 \parbox[t]{2mm}{\multirow{2}{*}{\rotatebox[origin=c]{90}{Dev}}} &
   CoNLL 2003&962&1,311&42\%&58\%\\[0.5pt]
 & OntoNotes5&1,159&1,524&43\%&57\%\\[0.5pt]
 \midrule
 \parbox[t]{2mm}{\multirow{2}{*}{\rotatebox[origin=c]{90}{Test}}} & 
   CoNLL 2003&879&1,228&42\%&58\%\\[0.5pt]
 & OntoNotes5&828&1,068&44\%&56\%\\[0.5pt]
 \bottomrule
\end{tabular}
\caption{Percentage of female and male names from the census data appearing in CoNLL 2003 and OntoNotes datasets with their corresponding counts. Both datasets fail to reflect the variety of female names.}
\label{data_stats_table}
\end{table}

In this work we not only performed a historical analysis of named entity recognition systems and showed the existence of bias, but we also introduced a benchmark that can help future models evaluate the extent of gender bias in their systems. We then performed a cross version analysis of models and showed that model updates can sometimes amplify the existing bias in previous versions. We also analyzed some datasets widely used in current state-of-the-art models and showed the existence of bias in these datasets as well which can directly affect the biased performance of models. Named entity recognition systems are extensively used in different downstream tasks and having biased NER systems can have implications beyond just the NER task. We believe that using our benchmark for evaluation of future named entity recognition systems can help mitigate the gender bias issue in these applications.

This work identifies an important problem with the current state-of-the-art in named entity recognition. Nevertheless, this measure is a glimpse into the many possible biases that NER may contain, and there are some key limitations that we plan to address in future work. First, the nine templates used to test the models are not necessarily representative of real-world text. There is a limitless supply of sentences that could be fed to the model. Moving forward, we seek to generate a sentence corpora that is based on real-world text. Second, our approach is based upon names taken from United States census data. This work can be extended to different languages to demonstrate the biases they pertain.

Through our analysis, we provide some suggestions for avoiding gender bias in NER systems, as listed below:
\begin{enumerate}
    \item Using benchmarks designed for evaluation of bias in future named entity recognition systems can help mitigate the gender bias issue in these applications. 
    \item Using contextual-based models as shown in our results can help reduce the error rates.
    \item We encourage introduction of new models to overcome the observed gender bias in these systems. As future work, these benchmarks can be used as loss functions to train the NER system. 
    \item A collection of new datasets reflecting reality, such as following name distributions observed in the real-world, would be helpful to the community to build models that avoid these biases. As a step towards this, we will provide our code and data upon acceptance.
\end{enumerate}

\section{Acknowledgments}
This material is based upon work supported by the Defense Advanced Research Projects Agency (DARPA) under Agreement No. HR0011890019. We would want to thank Kai-Wei Chang and Jieyu Zhao for their constructive feedbacks.

\bibliographystyle{aaai}
\bibliography{references}
\newpage
\begin{appendices}
\section{\\Weighted Results from Template \# 5}
\begin{figure}[h]
\includegraphics[width=\textwidth,height=\textwidth,trim=1cm 6cm 1cm 4cm,clip=true]{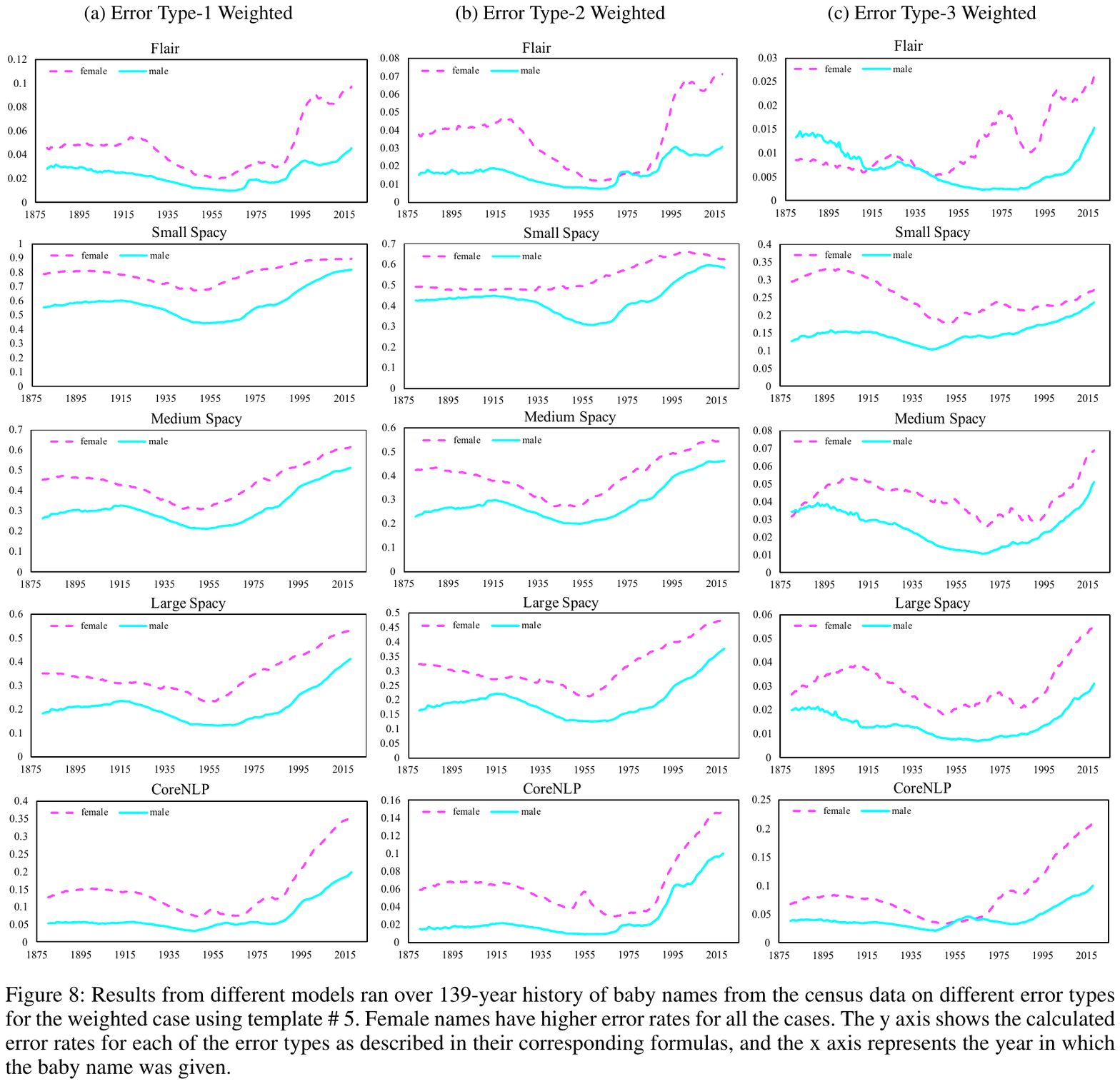}
\end{figure}

\clearpage
\section{\\Unweighted Results from Template \# 5}
\begin{figure}[h]
\centering
\includegraphics[width=\textwidth,height=\textwidth,trim=1cm 6cm 1cm 4cm,clip=true]{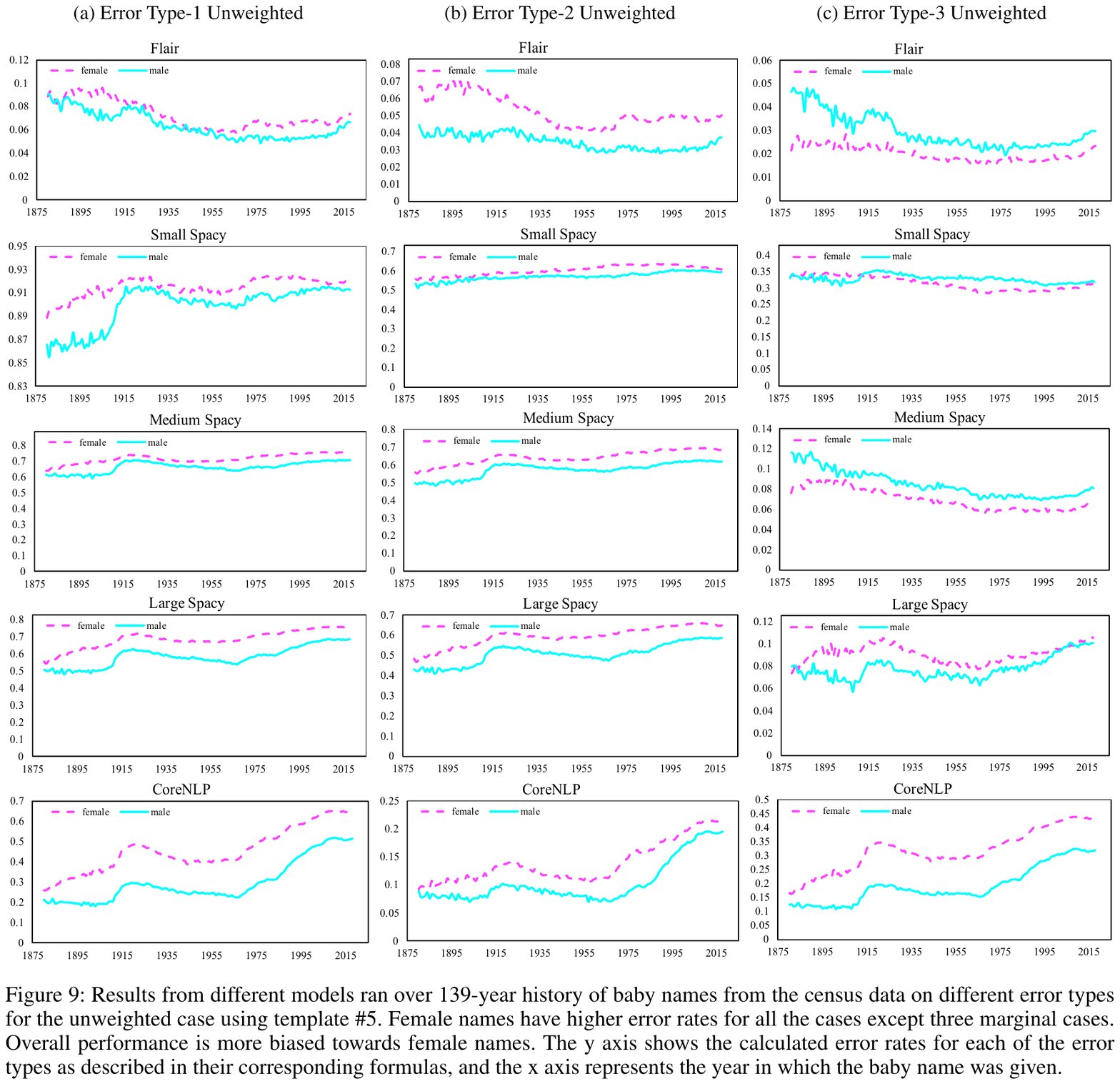}
\end{figure}
\end{appendices}

\end{document}